\documentclass[11pt]{amsart}  
\usepackage{amssymb}
\usepackage{amsmath}

\begin{document}

{\normalsize St. Petersburg Math. J. Vol. 17 (2006), No. 3, Pages 409-433. 
\footnote{In this version some translation errors have been corrected.}

\vskip 2.0cm

\title[Absence of eigenvalues for the periodic Dirac operator]{Absence of 
eigenvalues for the generalized two-dimensional periodic Dirac operator}
\author{L.I.{\,}Danilov}
\date{}

\address{Physical-Technical Institute, Ural Branch of the Russian Academy of Sciences,
Kirov Street 132, Izhevsk 426000, Russia}
\email{danilov@otf.pti.udm.ru}



\begin{abstract}
A generalized two-dimensional periodic Dirac operator is considered,
with $L^{\infty}$-matrix-valued coefficients of the first order derivatives
and with complex matrix-valued potential. It is proved that if the matrix-valued 
potential has zero bound relative to the free Dirac operator, then the spectrum
of the operator in question contains no eigenvalues.
\end{abstract}

\maketitle

\section*{Introduction}

Let ${\mathcal M}_2$ be the space of complex $(2\times 2)$-matrices,
$\widehat I\in {\mathcal M}_2$ the unit matrix, and
$$
\widehat \sigma _1=\left( \begin{matrix}  0 & 1\\ 1 & 0 \end{matrix} \right) 
\, ,\
\widehat \sigma _2=\left( \begin{matrix}  0 & -i\\ i & 0 \end{matrix} \right)
\, ,\
\widehat \sigma _3=\left( \begin{matrix}  1 & 0\\ 0 & -1 \end{matrix} \right)
$$
the Pauli matrices. Consider the generalized Dirac operator
$$
\widehat {\mathcal D}=\sum\limits
_{j=1}^2(h_{j1}\widehat {\sigma }_1+h_{j2}\widehat {\sigma }_2)\biggl( -i\,
\frac {\partial}{\partial x_j}\biggr) \, ,
$$
acting in $L^2({\bf R}^2;{\bf C}^2)$, with the domain 
$D(\widehat {\mathcal D})=H^1({\bf R}^2;{\bf C}^2)$. The functions $h_{jl}\in 
L^{\infty}({\bf R}^2;{\bf R})$ are assumed to be periodic with a (common) period
lattice $\Lambda \subset {\bf R}^2$, and $0<\varepsilon \leq h_{11}(x)h_{22}
(x)-h_{12}(x)h_{21}(x)$ for a.e. $x\in {\bf R}^2$. We denote by 
${\mathbb L}_{\Lambda}({\bf R}^2)$ the set of all $\Lambda $-periodic functions 
$W\in L^2_{\rm loc}({\bf R}^2;{\bf C})$ such that $W\varphi \in 
L^2({\bf R}^2)$ for any $\varphi \in H^1({\bf R}^2)$, and for every $\varepsilon >0$ 
there exists a number $C_{\varepsilon}(W)\geq 0$ satisfying 
$$
\| W\varphi \| _{L^2({\bf R}^2)}\leq \varepsilon \, \| \nabla \varphi \| 
_{L^2({\bf R}^2;{\bf C}^2)}+C_{\varepsilon}(W)\| \varphi \| _{L^2({\bf R}^2)}
$$
for all $\varphi \in H^1({\bf R}^2)$. If $V^{(l)}\in {\mathbb L}_{\Lambda}({\bf R}^2)$, 
$l=0,{\,}1,{\,}2,{\,}3$, then 
$$
\widehat {\mathcal D}+\widehat V=\widehat {\mathcal D}+V^{(0)}\widehat I+
\sum\limits_{l=1}^3V^{(l)}\widehat \sigma _l  \eqno (0.1)
$$
is a closed operator in $L^2({\bf R}^2;{\bf C}^2)$ with domain 
$D(\widehat {\mathcal D}+\widehat V)=D(\widehat {\mathcal D})=H^1({\bf R}^2;{\bf 
C}^2)$ (a $\Lambda $-periodic matrix-valued potential $\widehat V$ has zero
bound relative to $\widehat {\mathcal D}$ if and only if $V^{(l)}\in {\mathbb L}
_{\Lambda}({\bf R}^2)$, $l=0,{\,}1,{\,}2,{\,}3$).

The following theorem is the main result of this paper.
\vskip 0.2cm

{\bf Theorem 0.1}. {\it Let $h_{jl}\in L^{\infty}({\bf R}^2;{\bf R})$, $j,{\,}l
=1,{\,}2$, be periodic functions with period lattice $\Lambda \subset {\bf R}^2$.
Suppose that there exists $\varepsilon >0$ such that $\varepsilon \leq h_{11}(x)
h_{22}(x)-h_{12}(x)h_{21}(x)$ for a.e. $x\in {\bf R}^2$. If
$V^{(l)}\in {\mathbb L}_{\Lambda}({\bf R}^2)$, $l=0,{\,}1,{\,}2,{\,}3$, then 
the operator (0.1) has no egenvalues.}
\vskip 0.2cm

If under the conditions of Theorem 0.1, the operator (0.1) is selfadjoint, 
then its spectrum is absolutely continuous. For the selfadjoint periodic
elliptic differential operator, the absolute continuity of the spectrum is a
consequence of the absence of eigenvalues (of infinite multiplicity); see [1] 
(and also [2]). This fact is of general nature and is also valid for the
generalized Dirac operator $\widehat {\mathcal D}+\widehat V$.

The first results about the absence of eigenvalues in the spectrum of the
periodic Dirac operator were obtained in [3] -- [5]. For $n\geq 2$,
consider the $n$-dimensional periodic (with period lattice $\Lambda \subset 
{\bf R}^n$) Dirac operator
$$
\sum\limits_{j=1}^n\widehat \alpha _j\bigl( -i\, \frac {\partial}{\partial x_j}-
A_j\bigr) +\widehat V+\widehat V_0\, ,\ x\in {\bf R}^n\, , \eqno (0.2)
$$
where
$$
\widehat V=V\widehat I\, ,\ \widehat V_0=m\widehat \alpha
_{n+1}\, ,\ m\in {\bf R}\, ,  \eqno (0.3)
$$
the $\widehat \alpha _j\, $, $j=1,\dots ,n+1$, are Hermitian $(M\times M)$-matrices
satisfying the anticommutation relations $\widehat \alpha _j\widehat 
\alpha _l+\widehat \alpha _l\widehat \alpha _j=2\delta _{jl}\widehat I$, $\delta
_{jl}$ is the Kronecker symbol, and $\widehat I$ is the unit $(M\times M)$-matrix
($M\in 2{\bf N}$). The components $A_j$ of the vector-valued (magnetic) potential
and the scalar (electric) potential $V$ are real-valued periodic functions with
period lattice $\Lambda \subset {\bf R}^n$; let $K$ be the standard fundamental
domain of $\Lambda$. In [5,{\,}6], the absolute continuity of the spectrum of the
operator (0.2), (0.3) was proved for all $n\geq 2$ under the conditions $V\in 
C({\bf R}^n)$, $A\in L^{\infty}({\bf R}^n;{\bf R}^n)$, and
$$
\| \, |A|\, \| _{L^{\infty}({\bf R}^n)}<\max\limits_{\gamma \, \in \, \Lambda
\backslash \{ 0\} }\, \frac {\pi}{|\gamma |}  \eqno (0.4)
$$
($|x|$ is the length of the vector $x\in {\bf R}^n$). In subsequent papers, the
restriction on the periodic scalar potential $V$ has been relaxed. The spectrum
of the operator (0.2), (0.3) is absolutely continuous if (at least) one of the
following conditions is satisfied:

1) $n=2$, $V\in L^q(K)$, $q>2$, and the vector-valued potential $A\in L^{\infty}
({\bf R}^2;{\bf R}^2)$ satisfies (0.4) (see [7]);

2) $n\geq 3$, $A\equiv 0$, and $\sum\limits_N|V_N|^p<+\infty $, where the $V_N$ 
are the Fourier coefficients of $V$, $p\in [1,q_n(q_n-1)^{-1})$, and the $q_n>n$ 
are some numbers, the smallest (found before) values of which were presented 
for $n\geq 4$ in [8] (the numbers $q_n$ are found as the largest roots of the
algebraic equations
$$
q^4-(3n^2-4n-1)q^3+2(4n^2-6n-3)q^2-(9n^2-16n-4)q-4n(n-2)=0\, ,
$$
$q_3\simeq 11.645$, $n^{-2}q_n\to 3$ as $n\to +\infty $; for the first time
the above condition on the Fourier coefficients $V_N$ appeared in [4] for $n=3$);

3) $n=3$, $V\in L^q(K)$, $q>3$, and the vector-valued potential $A\in L^{\infty}
({\bf R}^3;{\bf R}^3)$ satisfies (0.4) (see [9]);

4) $n\geq 2$, $V\in L^2(K)$, $A\in L^{\infty}({\bf R}^n;{\bf R}^n)$, there exists
a vector $\gamma \in \Lambda \backslash \{ 0\} $ such that $\| \, |A|\, \| 
_{L^{\infty}({\bf R}^n)}<\pi |\gamma |^{-1}$ and the map
$$
{\bf R}^n\ni x\to \{ [0,1]\ni t\to V(x+t\gamma )\} \in L^2([0,1])
$$
is continuous; see [10]. (In particular, under an appropriate choice of $\gamma 
\in \Lambda \backslash \{ 0\} $, the latter condition is satisfied if $V$ is an
arbitrary piecewise continuous scalar potential with piecewise analytic
discontinuity surfaces; such potentials were considered in [3].) 

Some other conditions on the scalar potential $V$ and a small vector-valued
potential $A$ can be found in [10].

The periodic Dirac operator with a nonsmall vector-valued potential $A$
was studied in [11] -- [13]. In [12], the absolute continuity of the
spectrum of the operator (0.2) was proved under the conditions $\widehat V
=V\widehat I$, $\widehat V_0=V_0\widehat \alpha _{n+1}\, $, where $V,V_0\in 
L^q(K)$, $A\in L^q(K;{\bf R}^2)$, $q>2$, for $n=2$, and under the conditions 
$V,V_0\in C({\bf R}^n;{\bf R})$, $A\in C^{2n+3}({\bf R}^n;{\bf R}^n)$ for 
$n\geq 3$. For $n=2$, the proof is based on the results of the papers [14,{\,}15], 
where the periodic Schr\" odinger operator 
$$
\sum\limits_{j=1}^2\bigl( -i\, \frac {\partial}{\partial x_j}-A_j\bigr) ^2+V\, ,\
x\in {\bf R}^2\, .  \eqno (0.5)
$$
was treated. For the operator (0.5), the absolute continuity of the spectrum
was proved first for $V\in L^2_{\rm loc}({\bf R}^2;{\bf R})$, $A\in C({\bf R}^2;
{\bf R}^2)$ in [14], and then in [15] for the more general case where
$V\in L^q_{\rm loc}({\bf R}^2;{\bf R})$, $A\in L^{2q}_{\rm loc}({\bf R}^2;
{\bf R}^2)$, $q>1$ . For the periodic Dirac operator (0.2) with $n=2$, a similar 
result (as in [12]) was obtained in [11] (it was assumed, however, that $V_0\equiv 
m={\rm const}$, but the proof carries over to functions $V_0\in L^q(K)$, $q>2$,
without essential modifications). The methods used in [11] were the same as in [7]. 
More general conditions on $V$, $V_0$, and $A$ (for $n=2$) were obtained in [16]: 
it suffices to require that the functions $V^2\ln (1+|V|)$, $V_0^2\ln (1+|V_0|)$,
and $|A|^2\ln ^q(1+|A|)$ belong to $L^1(K)$ for some $q>1$. For $n\geq 3$, the
results of [12] were based on Sobolev's paper [17], where the absolute
continuity of the spectrum was proved for the Schr\" odinger operator with a
periodic vector-valued potential $A\in C^{2n+3}({\bf R}^n;{\bf R}^n)$. The latter
condition was relaxed by Kuchment and Levendorski$\breve {\mathrm i}$ in [2] 
(and also by Sobolev; see the remark at the end of the survey [18]): it suffices 
to require that $A\in H^q_{\rm loc}({\bf R}^n;{\bf R}^n)$, $2q>3n-2$, which makes it
possible to relax accordingly the constraint on the vector-valued potential
$A$ also for the periodic Dirac operator (see [12,{\,}18]).

Let ${\mathfrak M}_h$, $h>0$, be the set of all even Borel signed measures
(charges) $\mu$ on ${\bf R}$ (of finite total variation) for which
$\int_{{\bf R}}e^{\, ipt}d\mu (t)=1$ for every $p\in (-h,h)$. In 
[13], it was proved that the spectrum of the periodic (with period lattice 
$\Lambda \subset {\bf R}^n$) Dirac operator (0.2) is absolutely continuous for 
$n\geq 3$ if the following conditions are fulfilled:

1) the $(M\times M)$-matrix-valued functions $\widehat V$ and $\widehat  
V_0$ are Hermitian and continuous, and $\widehat V(x)\widehat \alpha _j=\widehat
\alpha _j\widehat V(x)$, $\widehat V_0(x)\widehat \alpha
_j=- \widehat \alpha _j\widehat V_0(x)$ for all $x\in {\bf R}^n$, $j=1,\dots ,n$;

2) $A\in C({\bf R}^n;{\bf R}^n)$ and there exist a vector $\gamma \in \Lambda
\backslash \{ 0\} $ and a measure $\mu \in {\mathfrak M}_h\, $, $h>0$, such that
for every $x\in {\bf R}^n$ and every unit vector $e\in {\bf R}^n$ with
$(e,\gamma )=0$ (where $(.,.)$ is the scalar product in ${\bf R}^n$) we have
$$
\biggl| \ \int\limits_{\bf R}d\mu (t)\int\limits_0^1A(x-\xi \gamma -te)\, d\xi -
A_0\, \biggr| <\frac {\pi}{|\gamma |}\, ,  \eqno (0.6)
$$
where $A_0=v^{-1}(K)\int_KA(x)\, d^{\, n}x$, $v(K)$ is the volume of the
fundamental domain $K$.

For a periodic vector-valued potential $A\in C({\bf R}^n;{\bf R}^n)$, condition
(0.6) is fulfilled (under an appropriate choice of $\gamma \in \Lambda \backslash 
\{ 0\} $ and measure $\mu \in {\mathfrak M}_h\, $, $h>0$) whenever $A\in H^q_{\rm loc}
({\bf R}^n;{\bf R}^n)$, $2q>n-2$, and also in the case where $\sum\limits_N|A_N|_{{\bf
C}^n}<+\infty $, where the $A_N$ are the Fourier coefficients of $A$ (see [13,{\,}19]).
The results of [13] were used in [20] in order to prove the absolute continuity of
the spectrum of the periodic Schr\" odinger operator
$$
\sum\limits_{j=1}^n\bigl( -i\, \frac {\partial}{\partial x_j}-A_j\bigr) ^2+V\, ,\
x\in {\bf R}^n\, ,\ n\geq 3\, ,
$$
with a vector-valued potential $A\in C^1({\bf R}^n;{\bf R}^n)$ satisfying condition
2) and with the scalar potential $V\in L^p_w(K;{\bf R})$ for which $t\, ({\rm
meas}\, \{ x\in K:|V(x)|>t\} )^{1/p}\to 0$ as $t\to +\infty $, where $p=n/2$ for
$3\leq n\leq 6$ and $p=n-3$ for $n\geq 7$, ${\rm meas}$ standing for Lebesgue
measure. The multidimensional periodic Schr\" odinger operator was studied in
many papers; the relevant facts and references can be found in [2,{\,}18] and
[21] -- [25].

Let ${\mathbb G}$ denote the set of all continuously differentiable and
monotone nonincreasing functions $g:(0,+\infty )\to (0,+\infty )$ (i.e., $g(t_1)
\geq g(t_2)$ for $0<t_1\leq t_2$) such that $\int_0^1(r{\,}g(r))^{-1}
dr < +\infty $ and $g(r/2)/g(r)\to 1$ as $r\to +0$. We write
$L^2\{ g,\Lambda \} $, $g\in {\mathbb G}$, for the Banach space of periodic
(with period lattice $\Lambda \subset {\bf R}^2$) functions $W\in L^2_{\rm loc}
({\bf R}^2;{\bf C})$ such that 
$$
\| W\| ^2_{L^2\{ g,\Lambda \} }=\sup\limits_{x\, \in \, {\bf R}^2}\ \int\limits
_{y\, :\, |x-y|\, \leq \, 1}g(|x-y|)\, |W(y)|^2d^{\, 2}y<+\infty \, .
$$
If $g\in {\mathbb G}$, then $g(r)(\ln r)^{-1}\to -\infty $ as $r\to +0$; therefore,
for any $W\in L^2\{ g,\Lambda \} $ the function $W^2$ belongs to the Kato class
$K_2$ (see [26]), which implies that $W\in {\mathbb L}_{\Lambda}({\bf R}^2)$.

For $n=2$, the operator $\widehat {\mathcal D}+\widehat V$ (see (0.1)) was considered
in [27] in the case where $h_{jl}\in C^{\infty}({\bf R}^2)$, $j,l=1,2$, $V^{(l)}\in 
C^{\infty}({\bf R}^2;{\bf R})$ for $l=1,2$, and $V^{(l)},\, \partial V^{(l)}/\partial 
x_j\in L^{\infty}({\bf R}^2)$ for $l=0,3$ and $j=1,2$. In [28], a special case of 
Theorem 0.1 was proved: under the same conditions on $h_{jl}\, $, it was assumed
that $\widehat V\in L^q_{\rm loc}({\bf R}^2;{\mathcal M}_2)$, $q>2$. The latter 
result was improved in [29] (and was announced in [30]): it suffices to require that
$V^{(0)},V^{(3)}\in {\mathbb L}_{\Lambda}({\bf R}^2)$ and $V^{(1)},V^{(2)}\in L^2\{ g,
\Lambda \} \subset {\mathbb L}_{\Lambda}({\bf R}^2)$ for some $g\in {\mathbb G}$.

The methods employed in the proof of Theorem 0.1 can also be used for the proof
of the absolute continuity of the spectrum of the two-dimensional periodic
Schr\" odinger operator
$$
\sum\limits_{j,l=1}^2\bigl( -i\, \frac {\partial}{\partial x_j}-A_j\bigr) G_{jl}
\bigl( -i\, \frac {\partial}{\partial x_l}-A_l\bigr) +V\, ,\
x\in {\bf R}^2\, ,  \eqno (0.7)
$$
with variable metric (see [29] -- [32]); here $G=(G_{jl})_{j,l=1,2}$ is a real 
symmetric positive definite matrix-valued function (a metric) with $G,{\,}G^{-1}\in 
L^{\infty}({\bf R}^2;{\mathcal M}_2)$. The functions $A$, $V$, and $G$ are periodic
with a common period lattice $\Lambda \subset {\bf R}^2$. For the first time, the
absolute continuity of the spectrum of the operator (0.7) was proved by Morame in [27]
under the conditions $A\in C^{\infty}({\bf R}^2;{\bf R}^2)$, $V\in L^{\infty}({\bf 
R}^2;{\bf R})$, and $G\in C^{\infty}({\bf R}^2;{\mathcal M}_2)$, ${\rm det}\, G\equiv 
1$. Later on, in the case where $G\in C^{\, m+\alpha }({\bf R}^2;{\mathcal M}_2)$, 
$m\in {\bf Z}_+\, $, $\alpha \in (0,1)$, Kuchment and Levendorski$\breve {\mathrm i}$
in [2] proved the existence of periodic isothermal coordinates $y(x)\in C^{\, 
m+1+\alpha }({\bf R}^2;{\bf R}^2)$ that reduce the matrix-valued function $G$ to a 
scalar form; the use of such coordinates allowed them to relax the restrictions on
$A$, $V$, and $G$ by reducing the problem to the case of a constant matrix $G$. 
The periodic isothermal coordinates were applied in a series of papers by Birman,
Suslina, and Shterenberg. In [33], the absolute continuity of the spectrum of the
operator (0.7) was proved for $G\in W^2_{2q,\, {\rm loc}}({\bf R}^2;{\mathcal M}_2)$, 
$A\in L^{2q}(K;{\bf R}^2)$, $q>1$, and $V=V_1+\sigma \delta _{\, \Sigma}\, $, where 
$V_1\in L^q(K;{\bf R})$, $\Sigma$ is a periodic system of piecewise smooth curves, 
$\delta _{\, \Sigma}$ is the $\delta $-function concentrated on $\Sigma$, and 
$\sigma \in L^q_{\rm loc}(\Sigma ;{\bf R})$. In the subsequent papers [34] -- [37], 
Shterenberg relaxed the conditions imposed on $A$, $V$, and $G$. The following
conditions were given in [36]:
$$
{\rm det}\, G\in H^1_{\rm loc}({\bf R}^2)\, ,\ \frac {\partial}{\partial x_j}\,
{\rm det}\, G\in {\mathbb L}_{\Lambda}({\bf R}^2)\, ,\ j=1,{\,}2\, ,
\eqno (0.8)
$$ $$
|A|^2\, \widetilde g(|A|)\in L^1_{\rm loc}({\bf R}^2)\, ,  \eqno (0.9)
$$
where $\widetilde g(t)=l(t)\doteq l^q_m(t)\prod\limits_{i=1}^{m-1}l_i(t)$,
$m\in {\bf N}$, $q>1$, $l_1(t)=1+\ln (1+t)$, $l_i(t)=1+\ln l_{i-1}(t)$, $i=2,
\dots ,m\, $, $t\geq 0$, and the scalar potential $V$ is defined as the distribution
$d\mu /d^{\, 2}x$, where $\mu$ is a periodic Borel signed measure satisfying some
additional conditions (which occur also in [35]); in this situation the closure of
the quadratic form ${\mathcal V}(\varphi ,\varphi )=\int\limits_{{\bf R}^2}|\varphi |
^2d\mu $, $\varphi \in C^{\infty}_0({\bf R}^2)$, may fail to be bounded relative to the 
form $\| \nabla \varphi \| _{L^2({\bf R}^2;{\bf C}^2)}^2\, $, $\varphi \in H^1({\bf 
R}^2)\subset L^2({\bf R}^2)$. Finally, in [37], the restriction (0.9) on the 
vector-valued potential $A$ was relaxed to $A_j\in {\mathbb L}_{\Lambda}({\bf R}^2)$, 
$j=1,2$. It should be mentioned that in [36,{\,}37] Shterenberg studied also a 
generalization of the operator (0.7) obtained by the incorporation of weight 
functions. The results pertaining to the absolute continuity of the spectrum of the 
periodic Schr\" odinger operator were applied to the study of the spectrum of the
Schr\" odinger operator in periodic waveguides (see [38,{\,}39] and also
[40,{\,}41]).

In [30] (without using periodic isothermal coordinates, and with the help of
results on the generalized two-dimensional periodic Dirac operator), it was proved 
that no eigenvalues are present in the spectrum of the periodic operator (0.7) if
the following conditions are fulfilled: $G$ satisfies (0.8), $A_j\in L^2\{ g,\Lambda 
\} $, $j=1,{\,}2$, for some $g\in {\mathbb G}$, and the scalar potential $V$ is
defined via a quadratic form ${\mathcal V}(\varphi ,\varphi )$, $\varphi \in H^1({\bf R}^2)$; 
this form has zero bound relative to the form $\| \nabla \varphi \| ^2_{L^2({\bf R}^2;
{\bf C}^2)}\, $, $\varphi \in H^1({\bf R}^2)\subset L^2({\bf R}^2)$, and is such that 
${\mathcal V}(\psi (.-\gamma ),\varphi (.-\gamma ))={\mathcal V}(\psi ,\varphi )$ for all  
$\psi ,\varphi \in H^1({\bf R}^2)$, $\gamma \in \Lambda $, and ${\mathcal V}(e^{\, 
i(k,x)}\psi ,e^{\, i(k,x)}\varphi )={\mathcal V}(\psi ,\varphi )$ for all $\psi ,\varphi \in 
H^1({\bf R}^2)$ and $k\in {\bf R}^2$ (if the form ${\mathcal V}(\varphi ,\varphi )$ is
Hermitian, then it may fail to coincide with $\int\limits_{{\bf R}^2}|\varphi |^2d\mu $
for $\varphi \in C^{\infty}_0({\bf R}^2)$, where $\mu$ is a periodic (locally finite) 
Borel signed measure [29]). In [30], the functions $A_j\, $, $j=1,2$, can be
complex-valued, and the form ${\mathcal V}$ is not assumed to be Hermitian.

If the function $[0,+\infty )\ni t\to \widetilde g(t)\in [0,+\infty )$ is monotone
nondecreasing and the function $(0,+\infty )\ni t\to \widetilde g(t^{-1})$ belongs
to ${\mathbb G}$ (in particular, this is true if $\widetilde g(.)=l(.)$), then for
every periodic vector-valued potential $A\in L^2_{\rm loc}({\bf R}^2;{\bf C}^2)$
satisfying (0.9) there exists $g\in {\mathbb G}$ such that $A_j\in L^2\{ g,\Lambda 
\} $, $j=1,{\,}2$ (see [31]). Therefore, the condition imposed in [30] on the
vector-valued potential $A$ is less restrictive than (0.9) (with $\widetilde g$
as indicated). In [32] (as well as in [37]), the constraint on $A$ was relaxed
up to $A_j\in {\mathbb L}_{\Lambda}({\bf R}^2)$, $j=1,2$. 

\section{Notation and the main statements}

Since the change $V^{(0)}-\lambda \to V^{(0)}$ reduces the operator $\widehat 
{\mathcal D}+\widehat V-\lambda \widehat I$, where $\widehat I$ is the identity
operator on $L^2({\bf R}^2;{\bf C}^2)$ and $\lambda \in {\bf C}$, to the operator 
$\widehat {\mathcal D}+\widehat V$, it follows that in the proof of Theorem 0.1 it
suffices to check the absence of the eigenvalue $\lambda =0$. Also, we may assume
that $\Lambda ={\bf Z}^2$ (an appropriate linear change of variables can be made,
preserving the form of the operator $\widehat {\mathcal D}+\widehat V$). Denote 
${\mathbb L}({\bf R}^2)\doteq {\mathbb L}_{{\bf Z}^2}({\bf R}^2)$, $K=[0,1)^2$. Let 
$0<q\leq p<+\infty $, let $F\geq 0$, and let $\Gamma (p,q,F)$ denote the set of 
ordered collections $\{ {\mathcal F},{\mathcal G},{\mathcal H}\} $ of
${\bf Z}^2$-periodic functions ${\mathcal F},{\mathcal G},{\mathcal H}$ in 
$L^{\infty}({\bf R}^2;{\bf R})$ such that $q\leq {\mathcal G}(x)\leq p$, $q\leq 
{\mathcal H}(x)\leq p$, and $|{\mathcal F}(x)|\leq F$ for a.e. $x\in {\bf R}^2$. 
We put $\Gamma =\bigcup\limits_{p,q,F}\Gamma (p,q,F)$. Multiplying the generalized 
Dirac operator (0.1) from the left by the (unitary) matrix-valued function
$$
(h^2_{21}(x)+h^2_{22}(x))^{-1/2}\, (h_{22}(x)\widehat I-ih_{21}(x)\widehat \sigma 
_3)\, ,\ x\in {\bf R}^2,
$$
we obtain the operator
$$
\widehat {\mathcal D}+\widehat V=({\mathcal G}\widehat \sigma _1+
{\mathcal F}\widehat \sigma _2)\bigl( -i\, \frac {\partial}{\partial x_1}\bigr)
+{\mathcal H}\widehat \sigma _2\bigl( -i\, \frac {\partial}{\partial x_2}\bigr)
+\widehat V\, ,  \eqno (1.1)
$$
for which $\{ {\mathcal F},{\mathcal G},{\mathcal H}\} \in \Gamma $ and the
(new) periodic matrix-valued potential $\widehat V$ satisfies the assumptions of 
Theorem 0.1. Therefore, Theorem 0.1 is a direct consequence of the next 
Theorem 1.1.
\vskip 0.2cm

{\bf Theorem 1.1}. {\it Suppose $\{ {\mathcal F},{\mathcal G},{\mathcal H}\} \in 
\Gamma $ and
$$
\widehat V=V^{(0)}\widehat I+\sum\limits_{l=1}^3V^{(l)}\widehat \sigma _l
$$
with $V^{(l)}\in {\mathbb L}({\bf R}^2)$, $l=0,{\,}1,{\,}2,{\,}3$. Then the 
generalized Dirac operator (1.1), acting in $L^2({\bf R}^2;{\bf C}^2)$ and defined
on the Sobolev class $H^1({\bf R}^2;{\bf C}^2)$, is invertible (i.e., $\lambda =0$
is not an eigenvalue of this operator).}
\vskip 0.2cm

For the proof of Theorem 1.1, we apply Thomas' method (originating from [42] and
used for checking the absence of eigenvalues in the spectrum of periodic elliptic
differential operators). With the help of this method, in this section we reduce
Theorem 1.1 to Theorem 1.2.

As a preliminary, we introduce some notation and several definitions. The Fourier
coefficients of a function $\varphi \in L^1(K;{\bf C}^d)$, $d=1,{\,}2$, will be
denoted by
$$
\varphi _N=\int\limits_K\varphi (x)\, e^{-2\pi i\, (N,\, x)}d^{\, 2}x\, ,\ N\in {\bf 
Z}^2.
$$
Let $\widetilde C(K)$, $\widetilde C^1(K)$, and $\widetilde H^1(K)$ be the spaces 
of functions $\varphi :K\to {\bf C}$, the ${\bf Z}^2$-periodic extensions of which
belong to $C({\bf R}^2)$, $C^1({\bf R}^2)$, and the Sobolev class $H^1_{\rm loc}
({\bf R}^2)$, respectively; by $\widetilde C_0(K)$, $\widetilde C^1_0(K)$, and 
$\widetilde H^1_0(K)$ we denote the corresponding subspaces of functions $\varphi$
such that $\varphi _0=\int\limits_K\varphi (x)\, d^{\, 2}x=0$; $\widetilde H^1(K;{\bf 
C}^2)=(\widetilde H^1(K))^2$. In what follows, we identify functions defined on
$K$ with their ${\bf Z}^2$-periodic extensions to ${\bf R}^2$. The norms and scalar
products in ${\bf C}^d$, $L^2({\bf R}^2;{\bf C}^d)$, and $L^2(K;{\bf C}^d)$, $d=
1,{\,}2$, are standard (as a rule, we do not indicate a particular space in the
notation for its norm and scalar product). The scalar products are assumed to be
linear in the second argument; $\nabla =(\partial /\partial x_1 \, ,\partial /
\partial x_2)$, and ${\rm meas}$ is Lebesgue measure on ${\bf R}^2$.

Let $\{ {\mathcal F} ,{\mathcal G} ,{\mathcal H}\} \in \Gamma (p,q,F)$. For all 
$k=(k_1,k_2)\in {\bf R}^2$ and all $\varkappa =(\varkappa _1,\varkappa _2)\in {\bf R}^2$, 
we introduce the operators
$$
\widehat {\mathcal D}(k+i\varkappa )=({\mathcal G}\widehat \sigma _1+{\mathcal
F}\widehat \sigma _2)(k_1+i\varkappa _1-i\, \frac {\partial}{\partial x_1})+
{\mathcal H}\widehat \sigma _2(k_2+i\varkappa _2-i\, \frac {\partial}{\partial 
x_2})\, ,
$$
acting in $L^2(K;{\bf C}^2)$, with $D(\widehat {\mathcal D}(k+i\varkappa ))=
\widetilde H^1(K;{\bf C}^2)$. Put
$$
\widehat d_{\pm}(k+i\varkappa )=({\mathcal G}\pm i{\mathcal F})(k_1+i\varkappa _1-i\, 
\frac {\partial}{\partial x_1})\pm i{\mathcal H}(k_2+i\varkappa _2-i\, \frac
{\partial}{\partial x_2})\, ,
$$
$D(\widehat d_{\pm}(k+i\varkappa ))=\widetilde H^1(K)\subset L^2(K)$, $\widehat 
d_{\pm}\doteq \widehat d_{\pm}(0)$;
$$
\widehat {\mathcal D}(k+i\varkappa )=\left( \begin{matrix} 0  &\widehat d_-(k+i
\varkappa ) \\ \widehat d_+(k+i\varkappa ) &0 \end{matrix} \right) \, .  \eqno (1.2)
$$
There exist numbers $c_1=c_1(p,q,F)>0$ and $c_2=c_2(p,q,F)\geq c_1$ such that for
all $k\in {\bf R}^2$ and all $\varphi \in \widetilde H^1(K)$ we have
$$
c_1\sum\limits_{j=1}^2\| \biggl( k_j-i\, \frac {\partial}{\partial x_j}\biggr)
\varphi \| ^2\leq \| \widehat d_{\pm}(k)\varphi \| ^2\leq c_2\sum\limits_{j=1}^2\| 
\biggl( k_j-i\, \frac {\partial}{\partial x_j}\biggr) \varphi   \| ^2  \eqno (1.3)
$$
(see, e.g., [29]).

Relations (1.2) and (1.3) imply that the operators $\widehat {\mathcal D}(k)$, 
$k\in {\bf R}^2$, are closed, and if $k\not\in 2\pi {\bf Z}^2$, then their range 
$R(\widehat {\mathcal D}(k))$ coincides with the space $L^2(K;{\bf C}^2)$,
${\rm ker}\, \widehat {\mathcal D}(k)=\{ 0\} $, and the inverse operators $\widehat 
{\mathcal D}^{-1}(k)$ are compact (see (1.2), (1.3), and the properties of
$\widehat d_{\pm}(k)$ presented in \S{\,}2).

The generalized Dirac operator $\widehat {\mathcal D}+\widehat V$ of the form (1.1) 
is unitarily equivalent to the direct integral
$$
{\int\limits_{2\pi K}}^{\oplus} \, (\widehat {\mathcal D}(k)+\widehat V)\, \frac 
{d^2k}{(2\pi )^2}\, ,  \eqno (1.4)
$$
acting in 
$$
{\int\limits_{2\pi K}}^{\oplus} \, L^2(K;{\bf C}^2)\, \frac {d^2k}{(2\pi )^2}
$$ 
(the vector $k=(k_1,k_2)\in 2\pi K\subset {\bf R}^2$ is called the {\it
quasimomentum}). The unitary equivalence mentioned above is established via the 
Gel$^{\prime }$fand transformation [43,{\,}44] (for periodic Dirac operators, see also 
[9,{\,}12]). The matrix-valued potential $\widehat V$, viewed as acting in $L^2(K;{\bf 
C}^2)$, has zero bound relative to the operators $\widehat {\mathcal D}(k)$, $k\in {\bf 
R}^2$; therefore, the operators $\widehat V\widehat {\mathcal D}^{-1}(k)$ are compact
for all $k\in {\bf R}^2\backslash 2\pi {\bf Z}^2$. Fix a vector $k^0\in 
{\bf R}^2\backslash 2\pi {\bf Z}^2$. Since
$$
(\widehat {\mathcal D}(k)+\widehat V)\widehat {\mathcal D}^{-1}(k^0)=
\widehat I+\widehat S(k)\, ,\ k\in {\bf R}^2,
$$
where $\widehat I$ is the identity operator in $L^2(K;{\bf C}^2)$, and the operator
$$
\widehat S(k)=(({\mathcal G}\widehat \sigma _1+{\mathcal F}\widehat \sigma _2)
(k_1-k^0_1)+{\mathcal H}\widehat \sigma _2(k_2-k^0_2)+\widehat V)
\widehat {\mathcal D}^{-1}(k^0)
$$
is compact, the representation of the operator (1.1) in the form of a direct
integral (1.4) and the analytic Fredholm theorem imply that if $\lambda =0$
is an eigenvalue of $\widehat {\mathcal D}+\widehat V$, then $\lambda =0$ is an
eigenvalue of each of the operators $\widehat {\mathcal D}(k+i\varkappa )+\widehat 
V$ (with the domain $D(\widehat {\mathcal D}(k+i\varkappa )+\widehat V)=\widetilde 
H^1(K;{\bf C}^2)\subset L^2(K;{\bf C}^2)$) for all $k+i\varkappa \in {\bf C}^2$
(see [1] and [44,{\,}{\S{\,}XIII.16}]). Consequently, for the proof of Theorem 1.1 
it suffices to find a complex vector $k+i\varkappa \in {\bf C}^2$ such that ${\rm ker}\, 
(\widehat {\mathcal D}(k+i\varkappa )+\widehat V)=\{ 0\} $. Thus, Theorem 1.1 is
implied by the following statement.
\vskip 0.2cm

{\bf Theorem 1.2}. {\it Suppose $\{ {\mathcal F}, {\mathcal G}, {\mathcal H}\} \in 
\Gamma (p,q,F)$,  
$$
\widehat V(.)=V^{(0)}(.)\widehat I+\sum\limits_{l=1}^3V^{(l)}(.)\widehat 
\sigma _l\, ,
$$
$V^{(l)}\in {\mathbb L}({\bf R}^2)$, $l=0,{\,}1,{\,}2,{\,}3$. Then there are vectors
$k^{\, \prime},\, \varkappa ^{\, \prime}\in {\bf R}^2$, and a unit vector $e=(e_1,e_2)
\in {\bf R}^2$ with $e_1>0$ such that for some numbers $\widetilde \mu >0$ as large
as we wish the following is true: for all $k\in {\bf R}^2$ with $k_1=\pi $ and all 
vector-valued functions $\varphi \in \widetilde H^1(K;{\bf C}^2)$ we have
$$
\| (\widehat {\mathcal D}(k+k^{\, \prime}+i(\widetilde \mu e+\varkappa ^{\, 
\prime}))+\widehat V)\varphi \| \geq e^{-c\, \widetilde \mu }\, \| \varphi \| \, ,
$$
where $c=c(p,q,F)>0$.}
\vskip 0.2cm

The proof of Theorem 1.2 is presented in {\S}{\,}5. It is based on Theorems 3.1 and
5.1, proved in {\S}{\,}3 and {\S}{\,}6, respectively. In {\S}{\,}2, for the
operators $\widehat d_{\pm}(k)\, $ we list the properties needed for what follows. 
In {\S}{\,}3 we prove that the Dirac operator $\widehat {\mathcal D}(0)+\widehat V$
with a matrix-valued potential $\widehat V$ of a special form is similar to the
Dirac operator $\widehat {\mathcal D}(k+i\varkappa )$ for some vectors $k,\, \varkappa \in 
{\bf R}^2$. In {\S}{\,}4 we collect the auxiliary statements to be used either in
the proof of Theorem 1.2, or (mainly) in the proof of Theorem 5.1. The estimates
proved in Theorem 5.1 for the Dirac operator $\widehat {\mathcal D}(k)+\widehat V$
with a matrix-valued potential of a special form are used in the proof of Theorem 1.2.
In {\S}{\,}6, Theorem 5.1 is deduced from Theorem 6.1, which is proved in the same
section. 

\section{Properties of the operators $\widehat d_{\pm}(k)$}

In this section we present the properties of $\widehat d_{\pm}(k)$, $k\in 
{\bf R}^2$, which we considered in detail in [29] and the proofs of which can be
found in [28,{\,}29]. In the case where $k=0$, we use the abbreviation $\widehat 
d_{\pm}\doteq \widehat d_{\pm}(0)$. The results of this section will be employed
substantially in what follows.

Estimates (1.3) imply that the operators $\widehat d_{\pm}(k)$, $k\in {\bf R}^2$,
are closed. If $k\notin 2\pi {\bf Z}^2$, then ${\rm ker}\, \widehat d_{\pm}(k)=\{
0\} $ and $R(\widehat d_{\pm}(k))=L^2(K)$; ${\rm ker}\, \widehat d_+={\rm ker}\, 
\widehat d_-$ is the one-dimensional subspace of constant functions in $L^2(K)$, 
and the subspaces $R(\widehat d_{\pm})$ are closed subspaces in $L^2(K)$ for which 
${\rm dim\, coker}\, \widehat d_{\pm}=1$. For all $\varphi \in \widetilde H^1(K)$ 
we have
$$
\overline {\widehat d_+\varphi }=-\widehat d_-\overline \varphi  \eqno (2.1)
$$
(the bar stands for complex conjugation). 

On the set $\Gamma $ we consider the metric
$$
\rho _{\infty}(\{ {\mathcal F}, {\mathcal G}, {\mathcal H}\} ,\{ {\mathcal F}^{\, 
\prime}, {\mathcal G}^{\, \prime}, {\mathcal H}^{\, \prime}\} )=
$$ $$
=\max \, \{ \| {\mathcal F}-{\mathcal F}^{\, \prime}\| _{L^{\infty}(K)}, \| 
{\mathcal G}-{\mathcal G}^{\, \prime}\| _{L^{\infty}(K)}, \| {\mathcal H}-
{\mathcal H}^{\, \prime}\| _{L^{\infty}(K)}\} \, .
$$
Let $\widehat P({\mathcal L})$ denote the orthogonal projection in $L^2(K)$ onto 
the subspace ${\mathcal L}$; the set of orthogonal projections is endowed with the
operator topology (induced by the operator norm). 
\vskip 0.2cm

{\bf Lemma 2.1} ([28]). {\it The functions $(\Gamma (p,q,F),\rho _{\infty})\ni \{ 
{\mathcal F}, {\mathcal G}, {\mathcal H}\} \to \widehat P({\rm coker}\, \widehat 
d_{\pm})$ are uniformly continuous.}
\vskip 0.2cm

This lemma and the convexity of the sets $\Gamma (p,q,F)$ imply that there
exist continuous functions 
$$
(\Gamma ,\rho _{\infty})\ni \{ {\mathcal F}, {\mathcal G}, {\mathcal H}\} \to 
\chi _{\pm}\in \{ \chi \in L^2(K):\| \chi \| =1\} \cap {\rm coker}\, \widehat
d_{\pm}\subset L^2(K)\, ,
$$
and we may assume that $\chi _-=\overline {\chi _+}\, $ (see (2.1)).
\vskip 0.2cm

{\bf Lemma 2.2} ([29]). {\it For any $\{ {\mathcal F}, {\mathcal G}, {\mathcal H}
\} \in \Gamma $ for a.e. $x\in K$ we have $\chi _+(x)\neq 0$.}
\vskip 0.2cm

{\bf Lemma 2.3} ([29]). {\it Let $\{ {\mathcal F}, {\mathcal G}, {\mathcal H}\} 
\in \Gamma (p,q,F)$, and let $g(.)$ be a positive, monotone nonincreasing, and
continuously differentiable function on $(0,1]$ such that $g(r/2)/g(r)\to 1$ 
as $r\to +0$. Then for all $x\in {\bf R}^2$ and all $\Phi \in \widetilde H^1(K)$
we have
$$
\int\limits_{y\, :\, |x-y|\, \leq \, 1}g(|x-y|)\, |\nabla \Phi (y)|_{{\bf C}^2}
^2\, d^{\, 2}y \leq c_3 \int\limits_{y\, :\, |x-y|\, \leq \, 1}g(|x-y|)\, 
|\widehat d_+ \Phi (y)|^2d^{\, 2}y\, ,
$$
where $c_3=c_3(p,q,F;g)>0$ (the integrals may take the value $+\infty $).}
\vskip 0.2cm

We denote by $\widetilde H^1_0\{ g,{\bf Z}^2\} $, $g\in {\mathbb G}$, the Banach
space of functions $\Phi \in \widetilde H^1_0(K)$ such that
$$
\| \Phi \| _{\widetilde H^1_0\{ g,{\bf Z}^2\} }\doteq \| \, |\nabla \Phi (.)|
\, \| _{L^2\{ g,{\bf Z}^2\} }<+\infty \, .
$$
By Lemma 2.3, $\widetilde H^1_0\{ g,{\bf Z}^2\} =\{ \Phi \in \widetilde H^1_0(K): 
\widehat d_+\Phi \in L^2\{ g,{\bf Z}^2\} \} $. Since $r^{\varepsilon}g(r)\to 0$ as 
$r\to +0$ for any $\varepsilon >0$, the space $\widetilde C^1_0(K)$ is embedded 
continuously in $\widetilde H^1_0\{ g,{\bf Z}^2\} $. On the other hand, 
$\widetilde H^1_0\{ g,{\bf Z}^2\} \subset \widetilde C_0(K)$, and for all 
$\Phi \in \widetilde H^1_0\{ g,{\bf Z}^2\} $ we have
$$
\| \Phi \| _{L^{\infty}(K)}\leq c_4\, \| \Phi \| _{\widetilde H^1_0\{ g,{\bf Z}
^2\} }\, ,  \eqno (2.2)
$$
with $c_4=c_4(g)>0$ (see [29]). The following Lemma 2.4 is an immediate consequence
of Lemma 2.3 and estimate (2.2).
\vskip 0.2cm

{\bf Lemma 2.4} ([29]). {\it If $\{ {\mathcal F}, {\mathcal G}, {\mathcal H}\} 
\in \Gamma (p,q,F)$ and $g\in {\mathbb G}$, then there is a number $c_5=c_5(p,q,
F;g)>0$ such that for any $\Phi \in \widetilde H^1_0(K)$ with $\widehat d_+\Phi 
\in L^2\{ g,{\bf Z}^2\} $ we have $\Phi \in \widetilde C_0(K)$ and
$$
\| \Phi \| _{L^{\infty}(K)}\leq c_5\, \| \widehat d_+ \Phi \| _{L^2\{ g,{\bf Z} 
^2\} }\, .
$$}
\vskip 0.2cm

The next lemma follows from the definition of the set ${\mathbb L}({\bf R}^2)$ 
(see, e.g., [22,{\,}36]).
\vskip 0.2cm

{\bf Lemma 2.5}. {\it Suppose $W\in {\mathbb L}({\bf R}^2)$. Then for any 
$\varepsilon >0$ there exists a number $C_{\varepsilon}(W)\geq 0$ such that for all
$k\in {\bf R}^2$ and all $\varphi \in \widetilde H^1(K)$ we have $W\varphi \in L^2(K)$ and
$$
\| W\varphi \| _{L^2(K)}\leq \varepsilon \, \| (k-i\nabla )\varphi \| _{L^2(K;{\bf C}^2)}+
C_{\varepsilon }(W)\, \| \varphi \| _{L^2(K)}\, .
$$}

{\bf Theorem 2.1} ([29]). {\it Suppose $\{ {\mathcal F}, {\mathcal G}, {\mathcal H}\} 
\in \Gamma $, $g\in {\mathbb G}$. Then for any $\Phi \in \widetilde H^1_0(K)$ with 
$\widehat d_+\Phi \in L^2\{ g,{\bf Z}^2\} $, and for any $\psi \in \widetilde H^1(K)$,
we have $\Phi \in \widetilde C_0(K)$, $e^{\, i\Phi }\psi \in \widetilde H^1(K)$, 
$(\partial \Phi / \partial x_j)\psi \in L^2(K)$, $j=1,{\,}2$, and
$$
\widehat d_+(e^{\, i\Phi }\psi )=e^{\, i\Phi }(i(\widehat d_+\Phi )\psi +\widehat
d_+\psi )\, .  \eqno (2.3)
$$}

Theorem 2.1 is a consequence of Lemmas 2.3, 2.4, and 2.5. For the proof of identity
(2.3), we use the fact that the operator $\widehat d_+$ is closed. First,
identity (2.3) is established for $\psi \in \widetilde C^1(K)$, and then, with
the use of Lemma 2.5, in the general case, for $\psi \in \widetilde H^1(K)$.
\vskip 0.2cm

{\bf Theorem 2.2} ([28]). {\it Suppose $\{ {\mathcal F}, {\mathcal G}, {\mathcal H}\} 
\in \Gamma $. Then there exist unique real-valued functions $\Phi ,\, \Psi \in 
\widetilde H^1_0(K)$ and a vector $\widetilde \varkappa \in {\bf R}^2$ such that
$$
i\widehat d_+(\Phi -i\Psi )=-({\mathcal G}+i{\mathcal F})\widetilde {\varkappa}_1-
i{\mathcal H}(\widetilde {\varkappa}_2+i)\, .  \eqno (2.4)  
$$
Moreover, (2.4) implies that $\Phi ,\, \Psi \in \widetilde C_0(K)$ and $\widetilde 
{\varkappa}_1>0$.}
\vskip 0.2cm

{\bf Lemma 2.6} ([29]). {\it Suppose $\{ {\mathcal F},{\mathcal G},{\mathcal H}\} \in 
\Gamma $. Then $\chi _+=c_6\, ({\mathcal G}{\mathcal H})^{-1}(\widehat d_+\Psi -
{\mathcal H})$, where $\Psi \in \widetilde H^1_0(K)\cap \widetilde C(K)$ is the
function defined in Theorem 2.2, and $c_6=c_6\, ({\mathcal F},{\mathcal G},{\mathcal 
H})\in {\bf C}\backslash \{ 0\} $.}
\vskip 0.2cm

{\bf Lemma 2.7} ([29]). {\it Suppose $\{ {\mathcal F},{\mathcal G},{\mathcal H}\} \in 
\Gamma $ and $\Psi \in \widetilde H^1_0(K)$ is the function defined in Theorem 2.2.
Then 
$$
\biggl( \frac {\partial \Psi }{\partial x_1}\biggr) ^2+\biggl( \frac {\partial 
\Psi }{\partial x_2}-1\biggr) ^2>0
$$
for a.e. $x\in K$.}

Lemma 2.7 is a consequence of Lemmas 2.2 and 2.6.
\vskip 0.2cm

{\bf Lemma 2.8} ([29]). {\it Under the conditions of Lemma 2.7, for all $\lambda 
\in {\bf R}$ we have
$$
{\rm meas}\, \{ x\in K:\Psi (x)-x_2=\lambda \} =0\, .
$$}

Lemma 2.8 follows from Lemma 2.7, because $\Psi \in H^1_{\rm loc}({\bf R}^2)$ and
if $\Psi (x)-x_2=\lambda \ (= {\rm const})$ on some set $M\subseteq K$ with 
${\rm meas}\, M>0$, then $\partial \Psi / \partial x_1=0$ and $\partial \Psi / 
\partial x_2=1$ for a.e. $x\in M$, which contradicts Lemma 2.7. Lemma 2.8 will be
used in the proof of Theorem 1.2. Instead of Lemma 2.8, we could apply Theorem 2.3
stated below (this method of argument was chosen in [28] for the proof of the
absence of eigenvalues in the spectrum of the generalized two-dimensional
periodic Dirac operator $\widehat {\mathcal D}+\widehat V$ with matrix-valued
potential $\widehat V\in L^q_{\rm loc}({\bf R}^2;{\mathcal M}_2)$, $q>2$).
\vskip 0.2cm

{\bf Theorem 2.3} ([28]). {\it Suppose $\{ {\mathcal F},{\mathcal G},{\mathcal H}\} 
\in \Gamma $ and identity (2.4) is fulfilled for real-valued functions $\Phi ,\, 
\Psi \in \widetilde H^1_0(K)\cap \widetilde C(K)$ and a vector $\widetilde \varkappa = 
(\widetilde \varkappa _1\, ,\widetilde \varkappa _2)\in {\bf R}^2$. Then ${\bf R}^2\ni 
x\to {\mathcal Z}(x)=\Phi (x)-i\Psi (x)+\widetilde {\varkappa}_1x_1+(\widetilde 
{\varkappa}_2+i)x_2\in {\bf C}$ is a continuous bijective map (with continuous
inverse).}
\vskip 0.2cm

Under the conditions of Theorem 2.3, ${\mathcal Z}(.)$ is a periodic map: for any
$x\in {\bf R}^2$ and $n\in {\bf Z}^2$, we have ${\mathcal Z}(x+n)={\mathcal Z}(x)
+\widetilde {\varkappa}_1n_1+(\widetilde {\varkappa}_2+i)n_2$, and $\widetilde {\varkappa}_1>0$.

\section{Similarity between the Dirac operator $\widehat {\mathcal D}(0)+\widehat V$
with a special matrix-valued potential and the Dirac operator $\widehat {\mathcal D}
(k+i\varkappa )$}

We put
$$
\widetilde H^1_0\{ {\mathbb G}\} =\bigcup\limits_{g\in {\mathbb G}}
\widetilde H^1_0\{ g,{\bf Z}^2\} \, .
$$
Lemma 2.3 and Theorem 2.1 imply that for any $\Phi \in \widetilde H^1_0\{ {\mathbb G}
\} $ the operators of multiplication by $e^{\, \Phi}$ and by $e^{\, \widehat {\sigma}
_3\Phi }$ act within the space $\widetilde H^1(K;{\bf C}^2)$.

{\bf Theorem 3.1}. {\it Suppose $\{ {\mathcal F}, {\mathcal G}, {\mathcal H} \} \in 
\Gamma (p,q,F)$ and $g\in {\mathbb G}$. Then for any two functions ${\mathcal C}_1,\, 
{\mathcal C}_2\in L^2\{ g,{\bf Z}^2\} $ there exist unique vectors $k,\, \varkappa \in 
{\bf R}^2$ and functions $\Phi ,\, \Psi \in \widetilde H^1_0\{ {\mathbb G}\} \subset 
\widetilde H^1_0(K)\cap \widetilde C(K)$ such that for some $\mu \in {\bf C}\backslash 
\{ 0\} $ (consequently, for all $\mu \in {\bf C}$) we have  
$$
e^{\, \mu \widehat \sigma _3\Psi }e^{-i\mu \Phi }\, \widehat {\mathcal D}(\mu (k+
i\varkappa ))\, e^{\, i\mu \Phi }e^{\, \mu \widehat \sigma _3\Psi }=\widehat 
{\mathcal D}(0)+\mu({\mathcal C}_1\widehat \sigma _1+{\mathcal C}_2\widehat 
\sigma _2)\, .  \eqno (3.1)
$$
Moreover, $\Phi ,\, \Psi \in \widetilde H^1_0\{ g,{\bf Z}^2\} $ and
$$
\max \{ \| \Phi \| _{L^{\infty}(K)}, \| \Psi \| _{L^{\infty}(K)}\} \leq
c^{\, \prime}_1\, (\| {\mathcal  C}_1\| _{L^2\{ g,{\bf Z}^2\} }+\| {\mathcal C}_2\|
_{L^2\{ g,{\bf Z}^2\} })\, ,  \eqno (3.2)
$$ $$
|k|^2+|\varkappa |^2\leq c^{\, \prime}_2\, (\| {\mathcal C}_1\|
^2_{L^2(K)}+\| {\mathcal C}_2\| ^2_{L^2(K)})\, ,  \eqno (3.3)
$$
where $c^{\, \prime}_1=c^{\, \prime}_1(p,q,F;g)>0$, $c^{\, \prime}_2=c^{\, \prime}
_2(p,q,F)>0$. If ${\mathcal C}_1\pm i\, {\mathcal C}_2\in R(\widehat d_{\pm})$, 
then $k=\varkappa =0$. If ${\mathcal C}_1$ and ${\mathcal C}_2$ are real-valued, 
then $\varkappa =0$ and $\Phi $ and $\Psi $ are also real-valued.}
\vskip 0.2cm

{\it Proof}. By Theorem 2.1 and relations (1.2) and (2.1), vectors $k,\, \varkappa 
\in {\bf R}^2$ and functions $\Phi ,\, \Psi \in \widetilde H^1_0\{ {\mathbb G}
\} \subset \widetilde H^1_0(K)\cap \widetilde C(K)$ satisfy (3.1) for some
$\mu \in {\bf C}\backslash \{ 0\} $ (and hence, for all $\mu \in {\bf C}$) 
if and only if, for both signs $+$ and $-{\, }$,
$$
i\widehat d_{\pm}\Phi _{\pm}={\mathcal C}^{\, \prime}_{\pm} \doteq {\mathcal
C}_{\pm}-(({\mathcal G}\pm i{\mathcal F})k_1\pm i{\mathcal H}k_2)-i(({\mathcal G}\pm i{\mathcal
F})\varkappa _1\pm i{\mathcal H}\varkappa _2)\, ,  \eqno (3.4)
$$
where $\Phi _{\pm}\doteq \Phi \mp i\Psi $, ${\mathcal C}_{\pm}\doteq {\mathcal C}_1
\pm i{\mathcal C}_2\in L^2\{ g,{\bf Z}^2\} \subset L^2(K)$ (multiplication by 
$e^{\, i\mu \Phi }$ and by $e^{\, \mu \Psi }$, $\mu \in {\bf C}$, acts within the
space $\widetilde H^1(K)$). We denote $(\chi _{\pm},{\mathcal G}\pm i{\mathcal F})
=\mu^{(1)}_{\pm}$, $(\chi _{\pm},\pm i{\mathcal H})=\mu ^{(2)}_{\pm}$. Since the
functions $\chi _{\pm}$ (with $\| \chi _{\pm}\| =1$) are chosen so that $\chi _-=
\overline {\chi _+}\, $, we have $\mu ^{(1)}_-=\overline {\mu ^{(1)}_+}$, $\mu 
^{(2)}_-=\overline {\mu ^{(2)}_+}$. Also, $|\mu ^{(1)}_{\pm}|\leq p+F$ and $|\mu 
^{(2)}_{\pm}|\leq p$. Since the subspaces $R(\widehat d_{\pm})$ are closed in 
$L^2(K)$ and ${\rm dim\, coker}\, \widehat d_{\pm}=1$, equations (3.4) can be
solved for $\Phi _{\pm}\in \widetilde H^1_0(K)$ and $k,{\,}\varkappa \in {\bf 
R}^2$ if and only if 
$$
(\chi _{\pm},{\mathcal C}_{\pm})=\mu ^{(1)}_{\pm}\, (k_1+i\varkappa _1)+\mu ^{(2)}
_{\pm}\, (k_2+i\varkappa _2)\, .  \eqno (3.5)
$$

{\bf Lemma 3.1}. {\it We have $|\, {\rm Im}\, \mu ^{(1)} _+\overline 
{\mu ^{(2)}_+}\, |\geq c_0=c_0(p,q,F)>0$.}
\vskip 0.2cm

{\it Proof}. Let ${\mathcal K}(x)=({\mathcal G}(x){\mathcal H}(x))
^{-1/2}$, $x\in K$. For any vector $\tau =(\tau _1,\tau _2)\in {\bf R}^2$ and any 
function $\Omega _+\in \widetilde H^1(K)$, we can write
$$
\| {\mathcal K}(i\widehat d_+\Omega _++({\mathcal G}+ i{\mathcal
F})\tau _1+ i{\mathcal H}\tau _2)\| ^2=
$$ $$
=\biggl\| {\mathcal K}{\mathcal G}(\tau _1+\frac {\partial \Omega _+}
{\partial x_1})\biggr\| ^2+\biggl\| {\mathcal K}\bigl( {\mathcal F}(\tau _1+\frac 
{\partial \Omega _+}{\partial x_1})+{\mathcal H}(\tau _2+\frac 
{\partial \Omega _+}{\partial x_2})\bigr) \biggr\| ^2\, ,
$$
whence
$$
\| i\widehat d_+\Omega _++({\mathcal G}+ i{\mathcal
F})\tau _1+ i{\mathcal H}\tau _2\| ^2\geq  \eqno (3.6)
$$ $$
\geq c_0\, \sum\limits_{j=1}^2\, \biggl\| \tau _j+\frac {\partial \Omega  
_+}{\partial x_j}\biggr\| ^2 =c_0\, \biggl( |\tau |^2+
\sum\limits_{j=1}^2\, \biggl\| \frac {\partial \Omega _+} {\partial 
x_j}\biggr\| ^2\, \biggr) \, ,
$$
where $c_0=c_0(p,q,F)>0$. Since the function $\Omega _+\in \widetilde H^1(K)$ 
in (3.6) is arbitrary, we obtain
$$
c_0\, |\tau |^2\leq \min\limits_{\Omega _+\in \widetilde H^1(K)}
\| i\widehat d_+\Omega _++({\mathcal G}+ i{\mathcal F})\, \tau _1+
i{\mathcal H}\, \tau _2\| ^2=
$$ $$
=|(\chi _+,{\mathcal G}+ i{\mathcal F})\, \tau _1+(\chi _+, i{\mathcal
H})\, \tau _2|^2=|\mu ^{(1)}_+\tau _1+\mu ^{(2)}_+\tau _2
|^2
$$
(in particular, this implies that $|\mu ^{(j)}_{\pm}|\geq \sqrt {c_0}$, $j=
1,{\,}2$). Consequently,
$$
c_0\leq \sqrt {c_0}\, |\mu ^{(2)}_+|\leq \min\limits_{t\in {\bf R}}|(\mu
^{(1)}_++\mu ^{(2)}_+t)\overline {\mu ^{(2)}_+}|=|\, {\rm Im}\, \mu ^{(1)}_+
\overline {\mu ^{(2)}_+}\, |\, .
$$
Lemma 3.1 is proved.  \hfill $\square$
\vskip 0.2cm

Since
$$
{\rm det}\, \left( \begin{matrix} \mu ^{(1)}_+  &\mu ^{(2)}_+\\  \mu ^{(1)}_-  
&\mu ^{(2)}_-\end{matrix} \right) = 2i\, {\rm Im}\, \mu ^{(1)} _+\overline {\mu 
^{(2)}_+}\, ,
$$
Lemma 3.1 shows that there exist unique vectors $k,\, \varkappa \in {\bf 
R}^2$ satisfying (3.5):
$$
k_1+i\varkappa _1=\bigl( 2i\, {\rm Im}\, \mu ^{(1)}_+\overline {\mu ^{(2)}_+}\bigr)
^{-1}(\mu ^{(2)}_-(\chi _+,{\mathcal C}_+)-\mu ^{(2)}_+(\chi _-,{\mathcal C}_-))
\, ,  \eqno (3.7)
$$ $$
k_2+i\varkappa _2=\bigl( 2i\, {\rm Im}\, \mu ^{(1)}_+\overline {\mu ^{(2)}_+}\bigr)
^{-1}(-\mu ^{(1)}_-(\chi _+,{\mathcal C}_+)+\mu ^{(1)}_+(\chi _-,{\mathcal C}_-))
\, .  \eqno (3.8)
$$
Relations (3.7) and (3.8) imply (3.3). For the vectors $k,\, \varkappa \in {\bf R}^2$ 
chosen as above, we have ${\mathcal C}_{\pm}^{\, \prime}\in R(\widehat d_{\pm})$, 
therefore, we can find (unique) functions $\Phi _{\pm}\in \widetilde H^1_0(K)$ such 
that $i\widehat d_{\pm}\Phi _{\pm}={\mathcal C}_{\pm}^{\, \prime}\, $. On the other 
hand, ${\mathcal C}_{\pm}^{\, \prime}\in L^2\{ g,{\bf Z}^2\} $, so that, by Lemmas 
2.3 and 2.4, $\Phi _{\pm}\in \widetilde H^1_0\{ g,{\bf Z}^2\} \subset \widetilde 
C_0(K)$ and
$$
\| \Phi _{\pm}\| _{L^{\infty}(K)}\leq c_5\, \| {\mathcal C}^{\, \prime}_{\pm}\|
_{L^2\{ g,{\bf Z}^2\} }\, .  \eqno (3.9)
$$
Also, we have $\Phi ,\, \Psi \in \widetilde H^1_0\{ g,{\bf Z}^2\} 
\subset \widetilde H^1_0(K)\cap \widetilde C(K)$, and (3.2) is implied by (3.3), 
(3.4), and (3.9). If ${\mathcal C}_{\pm}\in R(\widehat d_{\pm})$, then 
$(\chi _{\pm},{\mathcal C}_{\pm})=0$, whence $k=\varkappa =0$. If ${\mathcal 
C}_1$ and ${\mathcal C}_2$ are real-valued, then ${\mathcal C}_-=\overline {{\mathcal 
C}_+}$ and $(\chi _-,{\mathcal C}_-)=\overline {(\chi _+,{\mathcal C}_+)}$, and from
(3.7), (3.8) it follows that $\varkappa =0$ and $i\widehat d_{\pm}\Phi _{\pm}=
{\mathcal C}_{\pm}-(({\mathcal G}\pm i{\mathcal F})k_1\pm i{\mathcal H}k_2)$ (see
(3.4)). By complex conjugation and (2.1), we obtain $i\widehat d_{\pm}\overline 
{\Phi _{\mp}}={\mathcal C}_{\pm}-(({\mathcal G}\pm i{\mathcal F})k_1\pm i{\mathcal 
H}k_2)=i\widehat d_{\pm}\Phi _{\pm}$, whence $\widehat d_+({\rm Im}\, \Phi -i\, {\rm 
Im}\, \Psi )=0$. Consequently, $\Phi $ and $\Psi $ are real-valued. This proves
Theorem 3.1.            \hfill $\square$
\vskip 0.2cm

If under the assumptions of Theorem 3.1 we put ${\mathcal C}_1=i{\mathcal H}$, 
${\mathcal C}_2\equiv 0$, then ${\mathcal C}_1\, ,{\mathcal C}_2\in L^{\infty}(K)
\subset L^2\{ g,{\bf Z}^2\} $ for any $g\in {\mathbb G}$. Therefore, the next
theorem, which we need for what follows, is a consequence of Theorem 3.1. 
\vskip 0.2cm

{\bf Theorem 3.2}. {\it Suppose $\{ {\mathcal F},{\mathcal G},{\mathcal H}\} \in 
\Gamma (p,q,F)$. Then the following objects exist and are unique: a vector 
$\widetilde {\varkappa}\in {\bf R}^2$ and real-valued functions $\Phi ,\, \Psi \in 
\widetilde H^1_0\{ {\mathbb G}\} \subset \widetilde H^1_0(K)\cap \widetilde C(K)$  
such that for all $k,\, \varkappa \in {\bf R}^2$ and all $\mu \in {\bf R}$ we have
$$
e^{\, i\mu \widehat {\sigma}_3\Psi }e^{\, \mu \Phi }\widehat {\mathcal D}
(k+i\varkappa +i\mu \widetilde {\varkappa})e^{-\mu \Phi }e^{\, i\mu \widehat {\sigma}_3
\Psi }=\widehat {\mathcal D}(k+i\varkappa )+i\mu {\mathcal H}\widehat {\sigma}_1\, .
\eqno (3.10)
$$
Moreover, $\Phi ,\, \Psi \in \widetilde H^1_0\{ g,{\bf Z}^2\} $ for every $g\in 
{\mathbb G}$, and $\max \{ \| \Phi \| _{L^{\infty}(K)}\, ,\, \| \Psi \| _{L^{\infty}
(K)}\} \leq c^*_1$, $|\widetilde {\varkappa}|\leq c^*_2$, where $c^*_1=c^*_1\, (p,q,F)
>0$ and $c^*_2=c^*_2\, (p,q,F)>0$.}
\vskip 0.2cm

The functions $\Phi $, $\Psi $ and the vector $\widetilde \varkappa $ defined in
Theorem 3.2 coincide with the corresponding objects in Theorem 2.2, because they
satisfy condition (2.4), which is implied by (3.10).
\vskip 0.2cm

{\bf Lemma 3.2}. {\it For the vector $\widetilde \varkappa =(\widetilde \varkappa 
_1\, ,\widetilde \varkappa _2)\in {\bf R}^2$ defined in Theorem 3.2 we have
$\widetilde \varkappa _1\geq c_3^*=c_3^*(p,q,F)>0$.}
\vskip 0.2cm

{\it Proof}. Identity (2.4) yields
$$
\mu ^{(1)}_+\widetilde \varkappa _1+\mu ^{(2)}_+(\widetilde \varkappa _2+i)=0\, .
\eqno (3.11)
$$
Since $|\mu ^{(1)}_+|\leq p+F$ and $|\mu ^{(2)}_+|\geq \sqrt {c_0}$ (see the proof
of Lemma 3.1), from (3.11) we deduce that
$$
|\widetilde \varkappa _1|\geq \frac {\sqrt {c_0}}{p+F}\, |\, {\rm Im}\ \frac {\mu 
^{(1)}_+}{\mu ^{(2)}_+}\ \widetilde \varkappa _1\, |=\frac {\sqrt {c_0}}{p+F}\doteq
c_3^*\, ,
$$
and $\widetilde \varkappa _1>0$ by Theorem 2.2.           \hfill $\square$

\section{Auxiliary statements}

Let $k\in {\bf R}^2$, $\mu \in {\bf R}$. For every $N\in {\bf Z}^2$, we denote
$$
G^{\pm}_N(k;\mu )=\bigl( (k_1+2\pi N_1)^2+(
k_2+2\pi N_2\pm \mu )^2 \bigr) ^{1/2}\, ,
$$ $$
G_N(k;\mu )=\min \, \{ G^-_N(k;\mu ), G_N^+(k;\mu )\} 
$$
(here and in the sequel, we agree that the statements and formulas involving
$\pm$ and $\mp \, $ are understood independently for the upper and the lower
combination of signs). If $k_1=\pi $, then $G_N(k;\mu )\geq \pi $. For $\varphi \in 
\widetilde H^1(K)$, put
$$
\| \varphi \| _*=\biggl( \, \sum\limits_{N\in {\bf Z}^2}G^2_N(k;\mu )\, |\varphi
_N|^2 \biggr) ^{1/2}\, ,
$$ $$
\| \varphi \| _{*,\pm}=\biggl( \, \sum\limits_{N\in {\bf Z}^2}(G^{\pm}_N(k;\mu ))^2
\, |\varphi _N|^2 \biggr) ^{1/2}\, .
$$
For $a\geq 2\pi $, we introduce the finite sets
$$
T^{\pm}(a)=\{ N\in {\bf Z}^2: G^{\pm}_N(k;\mu )\leq a\} \, .
$$
In the above notation, the dependence on the vector $k\in {\bf R}^2$ and the
number $\mu \in {\bf R}$, which will be specified in advance, is not indicated
explicitly. Let $\# \, {\mathcal O}$ denote the number of elements of a finite
set ${\mathcal O}$. We have
$$
1\leq \# \, T^{\pm}(a) <6\pi a^2\, . \eqno (4.1)
$$
\vskip 0.2cm

{\bf Lemma 4.1}. {\it Suppose $\{ {\mathcal F}, {\mathcal G}, {\mathcal H}\} \in 
\Gamma (p,q,F)$. Then for all vectors $k\in {\bf R}^2$, all numbers
$\mu \in {\bf R}$, and all functions $\varphi \in \widetilde H^1(K)$ we have
$$
c_1\, \| \varphi \| ^2_{*,\pm}\leq \| (\widehat d_{\pm}(k)+i\mu {\mathcal H})\varphi \| 
^2\leq c_2\, \| \varphi \| ^2_{*,\pm}\, ,
$$
where $c_1=c_1(p,q,F)>0$ and $c_2=c_2(p,q,F)\geq c_1$.}
\vskip 0.2cm

This lemma is a consequence of estimates (1.3) (with the same constants $c_1$ 
and $c_2$).

For a set ${\mathcal O}\subseteq {\bf Z}^2$, denote ${\mathcal L}({\mathcal 
O})=\{ \psi \in L^2(K):\psi _N=0$ for $N\in {\bf Z}^2\backslash {\mathcal O}\}$, 
${\mathcal L}({\bf Z}^2)=L^2(K)$, ${\mathcal L}(\emptyset )=\{ 0\} $. Let $\widehat 
P^{\, {\mathcal O}}=\widehat P({\mathcal L}({\mathcal O}))$ be the orthogonal
projection in $L^2(K)$ that takes a function $\varphi \in L^2(K)$ to the function
$$
\widehat P^{\, {\mathcal O}}\varphi = \sum\limits_{N\in {\mathcal
O}}\varphi _Ne^{\, 2\pi i\, (N,\, x)}\, .
$$
\vskip 0.2cm

{\bf Lemma 4.2}. {\it If $W\in L^2(K)$, then for any finite set ${\mathcal O}\subset 
{\bf Z}^2$ the operator $W\widehat P^{\, \mathcal O}$ is bounded on $L^2(K)$, and
$$
\| W\widehat P^{\, \mathcal O}\| \leq f_W(\# \, {\mathcal O})\, ,
$$
where $f_W:{\bf Z}_+\to [0,+\infty )$ is a monotone nondecreasing function
satisfying $f_W(N)=o\, (\sqrt N)$ as $N\to +\infty $.}
\vskip 0.2cm

{\it Proof}. For $b\geq 0$, we introduce the following functions: 
$$
K\ni x\to W_b(x)=
\left\{
\begin{array}{ll}
W(x) & \text {if}\ |W(x)|>b, \\
0\ & \text {otherwise},
\end{array}
\right. \eqno (4.2)
$$
$\widetilde W_b(x)=W(x)-W_b(x)$, $x\in K$. If $\varphi \in L^2(K)$ and ${\mathcal 
O}\subset {\bf Z}^2$ is a finite set, then
$$
\| W_b\widehat P^{\, \mathcal O}\varphi \| = \biggl( \, \sum\limits_{N\in {\bf Z}^2}
\, \biggl| \, \sum\limits_{M\in {\bf Z}^2}(W_b)_M(\widehat P^{\, \mathcal O}\varphi )
_{N-M}\biggr| ^2 \, 
\biggr) ^{1/2}\leq
$$ $$
\leq \biggl( \, \sum\limits_{N\in {\bf Z}^2}\biggl( \, \sum\limits_{M\, :\, N-M
\in {\mathcal O}}|(W_b)_M|^2\biggr) \biggl( \, \sum\limits_{M\, :\, N-M\in 
{\mathcal O}}|\varphi _{N-M}|^2\biggr) \biggr) ^{1/2}\leq
$$ $$
\leq \biggl( \, \sum\limits_{M\in {\bf Z}^2}\biggl( \, \sum\limits_{N\, :\, N-M
\in {\mathcal O}}1\biggr) |(W_b)_M|^2\biggr) ^{1/2}\| \varphi \| = (\# \, {\mathcal O})
^{1/2}\, \| W_b \| _{L^2(K)}\| \varphi \| \, ,
$$
whence
$$
\| W\widehat P^{\, \mathcal O}\varphi \| \leq \| \widetilde W_b\widehat P^{\, 
\mathcal O}\varphi \| +\| W_b\widehat P^{\, \mathcal O}\varphi \| \leq (b+(\# \, {\mathcal 
O})^{1/2}\, \| W_b\| _{L^2(K)})\| \varphi \| \, .
$$
Put
$$
f_W(N)=\inf\limits_{b\geq 0}\ (b+\sqrt N\, \| W_b\| _{L^2(K)})\, ,\ N\in {\bf 
Z}_+={\bf N}\cup \{ 0\} \, .
$$
Then $\| W\widehat P^{\, \mathcal O}\| \leq f_W(\# \, {\mathcal O})$, the function 
$f_W$ is monotone nondecreasing, and for any $\varepsilon >0$ we can find a number 
$b(\varepsilon )>0$ such that $\| W_{b(\varepsilon )}\| _{L^2(K)}<\varepsilon $, 
so that $f_W(N)/{\sqrt N}\to 0$ as $N\to +\infty $.       \hfill $\square$
\vskip 0.2cm

Let $W\in {\mathbb L}({\bf R}^2)$; we put
$$
h_W(t)=\inf\limits_{\varepsilon >0}\ (\varepsilon +t^{-1}\, C_{\varepsilon}(W))\, ,\ 
t>0\, ,
$$
where $C_{\varepsilon}(W)$ is as in Lemma 2.5. The function $h_W$ is monotone
nonincreasing, and $h_W(t)\to 0$ as $t\to +\infty $. 
\vskip 0.2cm
 
{\bf Lemma 4.3}. {\it Suppose $W\in {\mathbb L}({\bf R}^2)$, $\mu 
\geq 4\pi $. Then for all $k\in {\bf R}^2$ with $k_1=\pi $ and all 
$\varphi \in {\mathcal L}(T^{\pm}(\mu /2))$ we have 
$$
\| W\varphi \| \leq c_7\, \| \varphi \| _{*,\pm} = c_7\, \| \varphi \| _*\, ,  
\eqno (4.3)
$$
where $c_7=c_7\, (W)>0$. If $2\pi \leq a\leq \mu /2$, then
$$
\| W\varphi \| \leq h_W(a)\| \varphi \| _*  \eqno (4.4)
$$
for all $k\in {\bf R}^2$ and all $\varphi \in {\mathcal L}(T^{\pm}(\mu /2)\backslash 
T^{\pm}(a))$.}

{\it Proof}. By Lemma 2.5 (with $\varepsilon =1$), for all $\mu
\in {\bf R}$, all $k\in {\bf R}^2$, and all $\varphi \in \widetilde
H^1(K)$ we have
$$
\| W\varphi \| \leq \| \varphi \| _{*,\, \pm }+C_1(W)\| \varphi \| \, .  \eqno (4.5)
$$
On the other hand, if $\varphi \in {\mathcal L}(T^{\pm}(\mu /2))$, $\mu \geq
4\pi $, then $\| \varphi \| _{*,\, \pm }=\| \varphi \| _*\, $, and $\| \varphi \|
_*\geq \pi \, \| \varphi \| $ whenever $k_1=\pi $. Therefore, (4.5) implies
(4.3) with $c_7=1+\pi ^{-1}C_1(W)$. Now, suppose that $2\pi \leq a\leq \mu 
/2$ and $\varphi \in {\mathcal L}(T^{\pm}(\mu /2)\backslash T^{\pm}(a))$. Then, for 
any $k\in {\bf R}^2$, we have $\| \varphi \| _{*,\, \pm }\geq a\, \| \varphi \| $,
and by Lemma 2.5 we obtain
$$
\| W\varphi \| \leq (\varepsilon +a^{-1}\, C_{\varepsilon}(W))\| \varphi \| _*
$$
for any $\varepsilon >0$, which yields (4.4).          \hfill $\square$
\vskip 0.2cm

{\bf Lemma 4.4}. {\it Suppose $W\in {\mathbb L}({\bf R}^2)$, $\mu 
\geq 4\pi $. Then for all $k\in {\bf R}^2$ and all $\varphi \in 
\widetilde H^1(K)\cap {\mathcal L}({\bf Z}^2\, \backslash (T^+(\mu /2)\cup T^-
(\mu /2)))$ we have 
$$
\| W\varphi \| \leq 3h_W(\mu )\, \| \varphi \| _*\, .
$$}

This lemma follows from Lemma 2.5, since, under the assumptions of Lemma 4.4,
we have $\| \varphi \| \leq 2\mu ^{-1}\| \varphi \| _*$ and $\| (k-i\nabla )
\varphi \| _{L^2(K;{\bf C}^2)}\leq 3\, \| \varphi \| _*\, $. 
\vskip 0.2cm

{\bf Lemma 4.5}. {\it Suppose $W\in L^2(K)$, $\mu >4\pi $, and $2\pi \leq
a<a^{\, \prime}\leq \mu /2$. Then for all $\varphi \in {\mathcal L}({\bf
Z}^2\backslash T^{\pm}(a^{\, \prime}))$ and all $\psi \in {\mathcal L}(T^{\pm}(a))$
we have
$$
|(\varphi ,W\psi )|\leq \sqrt {6\pi }\, a\, \biggl( \, \sum\limits_{N\in {\bf Z}^2\, 
:\, 2\pi |N|\, >\, a^{\, \prime}-a}|W_N|^2\, \biggr) ^{1/2}\, \| \varphi \| 
_{L^2(K)}\, \| \psi \| _{L^2(K)}\, .
$$}

{\it Proof}. Indeed,
$$
|(\varphi ,W\psi )|\leq \sum\limits_{M\in {\bf Z}^2}|\varphi _M|\, \sum\limits_{N\in 
{\bf Z}^2}|W_N\psi _{M-N}|\leq
$$ $$
\leq \| \varphi \| _{L^2(K)} \biggl( \, \sum\limits_{M\in {\bf Z}^2\backslash 
T^{\pm}(a^{\, \prime})} \biggl( \, \sum\limits_{N\in {\bf Z}^2}|W_N \psi _{M-N}|
\, \biggr) ^2 \, \biggr) ^{1/2}\leq
$$ $$
\leq \biggl( \, \sum\limits_{N\in {\bf Z}^2} \, \biggl( \, \sum\limits_{M\, :\, 
M\notin \, T^{\pm}(a^{\, \prime}),\, M-N\, \in \, T^{\pm}(a)}1\, \biggr) \, 
|W_N|^2\biggr) ^{1/2}\, \| \varphi \| _{L^2(K)} \, \| \psi \| _{L^2(K)} \leq
$$ $$
\leq \sqrt {6\pi }\, a\, \biggl( \, \sum\limits_{N\in {\bf Z}^2\, :\, 2\pi |N|\, >\, a^{\,
\prime}-a}|W_N|^2\biggr) ^{1/2}\, \| \varphi \| _{L^2(K)} \, \| \psi \|
_{L^2(K)} 
$$
(we have used estimate (4.1)). \hfill $\square$
\vskip 0.2cm

{\bf Lemma 4.6}. {\it For $W\in {\mathbb L}({\bf R}^2)$, let $W_b\, $, $b\geq 0$,
be the functions defined in (4.2). There exists a monotone nonincreasing function
$\widetilde h_W:[0,+\infty )\to [0,+\infty )$ such that $\widetilde h_W(t)\to
0$ as $t\to +\infty $ and for all $\mu \in {\bf R}$, all $k\in {\bf R}^2$ with
$k_1=\pi $, all $\varphi \in \widetilde H^1(K)$, and all $b\geq 0$ we have
$$
\| W_b\varphi \| \leq \widetilde h_W(b)\, \| \varphi \| _{*,\pm}\, .  \eqno (4.6)
$$}

{\it Proof}. By Lemma 2.5, for any $\varepsilon >0$ there exists a number 
$C_{\varepsilon}(W)\geq 0$ such that 
$$
\| W\psi \| \leq \varepsilon \, \| \psi \| _{*,\pm}+C_{\varepsilon}(W)\| \psi \| 
\eqno (4.7)
$$
for all $\mu \in {\bf R}$, $k\in {\bf R}^2$, and $\psi \in \widetilde H^1(K)$.
We define 
$$
\widetilde h_W(b)=\inf\limits_{\varepsilon >0}\, \min\limits_{a\geq 2\pi }\, \biggl(
\sqrt {\frac 6{\pi }}\, a\, \| W_b\| _{L^2(K)}+\varepsilon +a^{-1}C_{\varepsilon}(W)
\biggr) \, ,\ b\geq 0\, .
$$
Since the function $[0,+\infty )\ni b\to \| W_b\| _{L^2(K)}$ is monotone
nonincreasing, and $\| W_b\| _{L^2(K)}\to 0$ as $b\to +\infty $, the function 
$\widetilde h_W$ is also monotone nonincreasing, and $\widetilde h_W(b)\to 0$ as 
$b\to +\infty $. On the other hand, using (4.1) and (4.7), we see that if $\mu 
\in {\bf R}$, $k\in {\bf R}^2$, $k_1=\pi $, $\varphi \in \widetilde H^1(K)$ (in
which case $\pi \| \varphi \| \leq \| \varphi \| _{*,\pm}$), $b\geq 0$, $\varepsilon 
>0$ and $a\geq 2\pi $, then
$$
\| W_b\varphi \| \leq \| W_b\widehat P^{\, T^{\pm}(a)}\varphi \| +\| W_b\widehat P^{\,
{\bf Z}^2\backslash T^{\pm}(a)}\varphi \| \leq
$$ $$
\leq \| W_b\| _{L^2(K)}\| \widehat P^{\, T^{\pm}(a)}\varphi \| _{L^{\infty}(K)}+
\| W\widehat P^{\, {\bf Z}^2\backslash T^{\pm}(a)}\varphi \| \leq
$$ $$
\leq \sqrt {6\pi }\, a\, \| W_b\| _{L^2(K)}\| \widehat P^{\, T^{\pm}(a)}\varphi \| +
\varepsilon \, \| \widehat P^{\, {\bf Z}^2\backslash T^{\pm}(a)}\varphi \| _{*,\pm}+
C_{\varepsilon}(W)\| \widehat P^{\, {\bf Z}^2\backslash T^{\pm}(a)}\varphi \| \leq
$$ $$
\leq \biggl( \sqrt {\frac 6{\pi }}\, a\, \| W_b\| _{L^2(K)}+\varepsilon +a^{-1}C
_{\varepsilon}(W)\biggr) \, \| \varphi \| _{*,\pm}\, .
$$
These inequalities and the definition of $\widetilde h_W$ imply (4.6). 
\hfill $\square$

\section{Proof of Theorem 1.2}

For an arbitrary set ${\bf M}^{\, \prime}\subseteq {\bf N}$ we put
$$
{\mathcal Q}\, ({\bf M}^{\, \prime})=\overline {\lim\limits_{N\to +\infty }}\ \ 
\frac {\# \{ n\in {\bf M}^{\, \prime}:n\leq N\} }{N} \, .
$$
In this and the next sections we use the symbol $\sum\limits_{+,-}$ to denote
the sum of two terms obtained from expressions with indices $\pm$ and $\mp$ when
fixing the upper or the lower combination of signs.
\vskip 0.2cm

{\bf Theorem 5.1}. {\it Suppose $\{ {\mathcal F}, {\mathcal G}, {\mathcal H} \} \in 
\Gamma (p,q,F)$, $\widetilde V^{(l)}\in {\mathbb L}({\bf R}^2)$, $l=0,3$, and
$\Psi $ is a real-valued function of class $\widetilde C(K)$ such that 
$$
{\rm meas}\, \{ x\in K:\Psi (x)-x_2=\lambda \} =0   \eqno (5.1)
$$
for any $\lambda \in {\bf R}$. Then there exists a number $a_0=a_0(p,q,F;\widetilde 
V^{(0)},\widetilde V^{(3)})\geq 2\pi $ such that for any $a\geq a_0$ there is a set
${\bf M}\subset {\bf N}$, depending also on ${\mathcal F}$, ${\mathcal G}$, 
${\mathcal H}$, $\widetilde V^{(0)}$, $\widetilde V^{(3)}$, and $\Psi $, for 
which ${\mathcal Q}\, ({\bf N}\backslash {\bf M})=0$ and for all 
$\mu \in \pi {\bf M}$, all $k\in {\bf R}^2$ with $k_1=\pi $, and all 
$$
\varphi =\left( \begin{matrix} \varphi _+\\ \varphi _- \end{matrix} \right) \in \widetilde 
H^1(K;{\bf C}^2)\, ,  \eqno (5.2)
$$
we have the estimate
$$
\| (\widehat {\mathcal D}(k)+i\mu {\mathcal H}\widehat \sigma _1 +e^{\, 2i\mu 
\widehat \sigma _3\Psi }(\widetilde V^{(0)}\widehat I+\widetilde V^{(3)}
\widehat \sigma _3))\varphi \| ^2\geq  \eqno (5.3)
$$ $$
\geq \frac {c_1}6\, \sum\limits_{+,-}\| \widehat P^{\, T^{\pm}(a)}\varphi _{\pm}\|
^2_*+c_8 \sum\limits_{+,-}\| \widehat P^{\, {\bf Z}^2\backslash T^{\pm}(a)}\varphi 
_{\pm}\| ^2_{*,\pm}\, ,
$$
where $c_8=c_8(p,q,F;\widetilde V^{(0)},\widetilde V^{(3)})\in (0,\frac 16\, c_1] 
$.}
\vskip 0.2cm

The proof of this theorem is postponed until {\S}{\,}6.

We pass to the proof of Theorem 1.2, in which Theorems 3.1, 3.2, and 5.1 will play
an important part. For $l=1,2$ and $b\geq 0$ (as in (4.2)), we introduce the
functions
$$
{\bf R}^2\ni x\to V^{(l)}_b(x)=
\left\{
\begin{array}{ll}
V^{(l)}(x) & \text {if}\ |V^{(l)}(x)|>b, \\
0\ & \text {otherwise}.
\end{array}
\right. 
$$
Since $V^{(l)}\in {\mathbb L}({\bf R}^2)$, $l=1,2$, Lemma 4.6 allows us to choose
a number $b=b(c_1;V^{(1)},V^{(2)})\geq 0$ so that for all $\mu \in {\bf R}$, all 
$k\in {\bf R}^2$ with $k_1=\pi $, and all $\psi \in \widetilde H^1(K)$ we have
the inequalities
$$
\| V^{(l)}_b\psi \| ^2\leq \frac {c_1}{192}\, \| \psi \| ^2_{*,\pm}\, ,\ l=1,2\,
,  \eqno (5.4)
$$
for both signs $+$ and $-\, $. For $l=1,2$ we have $\| V^{(l)}-V^{(l)}_b\| 
_{L^{\infty}({\bf R}^2)}\leq b<+\infty $; therefore, by Theorem 3.1, there exist
vectors $k^{\, \prime}, {\varkappa}^{\, \prime}\in {\bf R}^2$ and functions $\Phi 
^{\, \prime}, \Psi ^{\, \prime}\in \widetilde H^1_0(K)\cap \widetilde C(K)$ with the
following properties: the operators of multiplication by the functions $e^{\, \pm 
i\Phi ^{\, \prime}}$ and by the matrix-valued functions $e^{\, \pm \widehat \sigma 
_3\Psi ^{\, \prime}}$ act within the space $\widetilde H^1(K;{\bf C}^2)$, for every 
$k,\varkappa \in {\bf R}^2$ we have
$$
e^{\, \widehat \sigma _3\Psi ^{\, \prime}}e^{-i\Phi ^{\, \prime}}\, (\widehat 
{\mathcal D}(k+k^{\, \prime}+i(\varkappa +\varkappa ^{\, \prime}))+\widehat V)\, e^{\, 
i\Phi ^{\, \prime}}e^{\, \widehat \sigma _3\Psi ^{\, \prime}}=  \eqno (5.5)
$$ $$
=\widehat {\mathcal D}(k+i\varkappa )+\widetilde V^{(0)}\widehat I+\sum\limits_{l=1}
^2V^{(l)}_b\widehat \sigma _l+\widetilde V^{(3)}\widehat \sigma _3\, ,
$$
where
$$
\widetilde V^{(0)}=V^{(0)} \cosh 2\Psi ^{\, \prime} +V^{(3)} \sinh 2\Psi ^{\, \prime}
\, ,\ \widetilde V^{(3)}=V^{(0)} \sinh 2\Psi ^{\, \prime} +V^{(3)} \cosh 2\Psi ^{\, 
\prime}\, ,
$$
and
$$
\max \{ \| \Phi ^{\, \prime}\| _{L^{\infty}(K)}, \| \Psi ^{\, \prime}\| 
_{L^{\infty}(K)}\} \leq c^{\, \prime \prime}_1\, b\, ,  \eqno (5.6)
$$
where $c^{\, \prime \prime}_1=c^{\, \prime \prime}_1(p,q,F)>0$. Inequality (5.6) 
shows that $\widetilde V^{(0)},\widetilde V^{(3)}\in {\mathbb L}({\bf R}^2)$. Let
$\Phi , \Psi \in \widetilde H^1_0(K)\cap \widetilde C(K)$ and $\widetilde \varkappa 
\in {\bf R}^2$ be the vector-valued functions and the vector defined in Theorem
3.2 (for the functions ${\mathcal F}$, ${\mathcal G}$, ${\mathcal H}$). 
Multiplications by $e^{\, \mu \Phi }$ and by $e^{\, i\mu \widehat 
{\sigma}_3\Psi }$, $\mu \in {\bf R}$, also act within the space
$\widetilde H^1(K;{\bf C}^2)$. From (3.10) we obtain
$$
e^{\, i\mu \widehat {\sigma}_3\Psi }e^{\, \mu \Phi }\, \bigl( \widehat {\mathcal 
D}(k+i\mu \widetilde {\varkappa})+\widetilde V^{(0)}\widehat I+\sum\limits_{l=1}
^2V^{(l)}_b\widehat \sigma _l+\widetilde V^{(3)}\widehat \sigma _3\bigr) \, 
e^{-\mu \Phi }e^{\, i\mu \widehat {\sigma}_3\Psi }=  \eqno (5.7)
$$ $$
=\widehat {\mathcal D}(k)+i\mu {\mathcal H}\widehat {\sigma}_1+\sum\limits_{l=1}
^2V^{(l)}_b\widehat \sigma _l+e^{\, 2i\mu \widehat {\sigma}_3\Psi }\, (\widetilde 
V^{(0)}\widehat I+\widetilde V^{(3)}\widehat \sigma _3)
$$
for all $\mu \in {\bf R}$ and all $k\in {\bf R}^2$. By Lemma 2.8, $\Psi $ 
satisfies (5.1). Let $a_0$ and $c_8$ be the numbers defined in Theorem 5.1 (for the 
functions $\widetilde V^{(0)}$, $\widetilde V^{(3)}$, and $\Psi $). We put 
$\varepsilon =\frac 1{16}\, \sqrt {2c_8}$ and choose a number $a\geq a_0$ so that 
$\varepsilon a\geq C_{\varepsilon}(V^{(l)})$, $l=1,2$, where $C_{\varepsilon}(.)$ is
as in (4.7). By Theorem 5.1, there exists a set ${\bf M}\subset {\bf N}$, depending 
on ${\mathcal F}$, ${\mathcal G}$, ${\mathcal H}$, on the matrix-valued potential 
$\widehat V$, and also on the choice of $b$ and $a$, such that ${\mathcal Q}({\bf N}
\backslash {\bf M})=0$ and estimate (5.3) is valid for all $\mu \in \pi {\bf M}$, all 
$k\in {\bf R}^2$ with $k_1=\pi $, and all vector-valued functions (5.2). Using (4.7), 
(5.4), and the estimates
$$
\| \widehat P^{\, {\bf Z}^2\backslash T^{\pm}(a)}\varphi _{\pm}\| \leq \frac 1a\
\| \widehat P^{\, {\bf Z}^2\backslash T^{\pm}(a)}\varphi _{\pm}\| _{*,\pm}\, ,
$$
we obtain the inequalities
$$
\| \sum\limits_{l=1}^2V^{(l)}_b\widehat \sigma _l\, \varphi \| ^2\leq
2\sum\limits_{l=1}^2\| V^{(l)}_b \varphi \| ^2\leq
$$ $$
\leq 4\sum\limits_{l=1}^2\sum\limits_{+,-}\| V^{(l)}_b\widehat P^{\, T^{\pm}(a)}
\varphi _{\pm}\| ^2+4\sum\limits_{l=1}^2\sum\limits_{+,-}\| V^{(l)}\widehat P^{\, 
{\bf Z}^2\backslash T^{\pm}(a)}\varphi _{\pm}\| ^2 \leq
$$ $$
\leq \frac {c_1}{24}\, \sum\limits_{+,-}\| \widehat P^{\, T^{\pm}(a)}\varphi _{\pm}
\| ^2_{*,\pm}+
$$ $$
+8\sum\limits_{l=1}^2\sum\limits_{+,-}\biggl( \varepsilon ^2\| \widehat 
P^{\, {\bf Z}^2\backslash T^{\pm}(a)}\varphi _{\pm}\| ^2_{*,\pm}+C_{\varepsilon}^2(V^
{(l)})\| \widehat P^{\, {\bf Z}^2\backslash T^{\pm}(a)}\varphi _{\pm}\| ^2\biggr)
\leq
$$ $$
\leq \frac {c_1}{24}\, \sum\limits_{+,-}\| \widehat P^{\, T^{\pm}(a)}\varphi _{\pm}
\| ^2_{*,\pm}+\frac {c_8}4\, \sum\limits_{+,-}\| \widehat P^{\, {\bf Z}^2
\backslash T^{\pm}(a)}\varphi _{\pm}\| ^2_{*,\pm}\, .
$$
Therefore, (5.3) implies that, again for all $\mu \in \pi {\bf M}$, all 
$k\in {\bf R}^2$ with $k_1=\pi $, and all $\varphi $ as in (5.2), 
$$
\| (\widehat {\mathcal D}(k)+i\mu {\mathcal H}\widehat {\sigma}_1+\sum\limits
_{l=1}^2V^{(l)}_b\widehat \sigma _l+e^{\, 2i\mu \widehat {\sigma}_3\Psi }\, 
(\widetilde V^{(0)}\widehat I+\widetilde V^{(3)}\widehat \sigma _3))\varphi \| ^2
\geq  \eqno (5.8)
$$ $$
\geq \frac 12\, \| (\widehat {\mathcal D}(k)+i\mu {\mathcal H}\widehat {\sigma}_1
+e^{\, 2i\mu \widehat {\sigma}_3\Psi }\, (\widetilde V^{(0)}\widehat I+\widetilde 
V^{(3)}\widehat \sigma _3))\varphi \| ^2- \| \sum\limits_{l=1}^2V^{(l)}_b\widehat 
\sigma _l\, \varphi \| ^2 \geq
$$ $$
\geq \frac {c_1}{24}\, \sum\limits_{+,-}\| \widehat P^{\, T^{\pm}(a)}\varphi _{\pm}
\| ^2_{*,\pm}+\frac {c_8}4\, \sum\limits_{+,-}\| \widehat P^{\, {\bf Z}^2
\backslash T^{\pm}(a)}\varphi _{\pm}\| ^2_{*,\pm}\geq
$$ $$
\geq \frac {c_8}4\, \sum\limits_{+,-}\| \varphi _{\pm}\| ^2_{*,\pm}\geq
\frac {c_8}4\, \sum\limits_{N\in {\bf Z}^2}G^2_N(k;\mu )|\varphi _N|^2\geq
\frac {\pi ^2}4\, c_8\, \| \varphi \| ^2\, .
$$
Now we use (5.5) with $\varkappa =\mu \widetilde \varkappa $ and also (5.6) -- (5.8) 
and the estimate $\| \Phi \| _{L^{\infty}(K)}\leq c_1^*$ (see Theorem 3.2) to obtain
the inequality
$$
\| (\widehat {\mathcal D}(k+k^{\, \prime}+i(\mu \widetilde \varkappa +\varkappa ^{\, 
\prime}))+\widehat V)\varphi \| \geq \frac {\pi}2\, \sqrt {c_8}\, e^{-4c_1^{\, 
\prime \prime}b}\, e^{-2c_1^*\mu }\, \| \varphi \| \, ,
$$
which is valid for all $\mu \in \pi {\bf M}$, all $k\in {\bf R}^2$ with $k_1=\pi $,
and all $\varphi \in \widetilde H^1(K;{\bf C}^2)$. To complete the proof of Theorem 
1.2 it remains to put $e=\widetilde \varkappa /|\widetilde \varkappa |$ and $c=
3c_1^*/c_3^*$ (see Lemma 3.2), and it suffices to choose numbers $\widetilde \mu 
\doteq |\widetilde \varkappa |\mu \in \pi |\widetilde \varkappa |\, {\bf M}$ for
which
$$
4c_1^{\, \prime \prime}\, b-\ln \bigl( \, \frac {\pi}2\, \sqrt {c_8}\, \bigr) 
\leq c_1^*\, \frac {\widetilde \mu }{|\widetilde \varkappa |}\, .
$$
Theorem 1.2 is proved.    \hfill $\square$
\vskip 0.2cm

{\it Remark}. Under the conditions of Theorem 1.2, if $V^{(l)}\in L^2\{ g,{\bf 
Z}^2\} $, $l=1,2$, for some $g\in {\mathbb G}$, then the proof of Theorem 1.2
simplifies, because Theorem 3.1 provides an identity similar to (5.5) but without
the term  $\sum\limits_{l=1}^2V^{(l)}_b\widehat \sigma _l$ on the right-hand side 
(see also [29]).

\section{Proof of Theorem 5.1}
 
Under the conditions of Theorem 5.1, denote $V^{\, (\pm )}=\widetilde V^{(0)}\pm 
\widetilde V^{(3)}$. Since 
$$
\widehat {\mathcal D}(k)+i\mu {\mathcal H}\widehat \sigma _1+e^{\, 2i\mu \widehat 
\sigma _3\Psi }\, (\widetilde V^{(0)}\widehat I+\widetilde V^{(3)}\widehat \sigma 
_3)=\left( \begin{matrix} e^{\, 2i\mu \Psi }\, V^{\, (+)} &\widehat
d_-(k)+i\mu {\mathcal H}\\ \widehat d_+(k)+i\mu {\mathcal H} &e^{-2i\mu \Psi }\, 
V^{\, (-)}\end{matrix} \right) \, ,
$$ 
Theorem 5.1 is equivalent to the following statement.
\vskip 0.2cm

{\bf Theorem 6.1}. {\it Suppose $\{ {\mathcal F}, {\mathcal G}, {\mathcal H} \} \in 
\Gamma (p,q,F)$, $V^{\, (\pm )}\in {\mathbb L}({\bf R}^2)$, and $\Psi $ is a
real-valued function of class $\widetilde C(K)$ satisfying (5.1) for all $\lambda 
\in {\bf R}$. Then there exists a number $a_0^{\, \prime}=a_0^{\, \prime}(p,q,F;V^{\, 
(+)},V^{\, (-)})\geq 2\pi $ with the following property: for every $a_1\geq a_0^{\, 
\prime}$ we can find a set ${\bf M}\subset {\bf N}$, depending also on ${\mathcal F}$,
${\mathcal G}$, ${\mathcal H}$, $V^{\, (+)}$, $V^{\, (-)}$, and $\Psi $, such that 
${\mathcal Q}\, ({\bf N}\backslash {\bf M})=0$ and for all $\mu \in \pi {\bf M}$, all 
$k\in {\bf R}^2$ with $k_1=\pi $, and all $\varphi _{\pm}\in \widetilde H^1(K)$ we
have
$$
\| (\widehat d_+(k)+i\mu {\mathcal H})\varphi _++e^{-2i\mu \Psi }\, V^{\, (-)}
\varphi _- \| ^2+\| (\widehat d_-(k)+i\mu {\mathcal H})\varphi _-+e^{\, 2i\mu \Psi }\, 
V^{\, (+)}\varphi _+ \| ^2\geq
$$ $$
\geq \frac {c_1}{6}\, \bigl( \| \widehat P^{\, T^+(a_1)}\varphi _+\| ^2_*+
\| \widehat P^{\, T^-(a_1)}\varphi _-\| ^2_*\bigr) +
$$ $$
+c^{\, \prime}_8\, \bigl( \| \widehat P^{\, {\bf Z}^2\backslash T^+(a_1)}\varphi _+
\| ^2_{*,+}+\| \widehat P^{\, {\bf Z}^2\backslash T^-(a_1)}\varphi _-\| ^2_{*,-}
\bigr) \, ,
$$
where $c_8^{\, \prime}=c_8^{\, \prime}(p,q,F;V^{\, (+)},V^{\,
(-)})\in (0,\frac 16\, c_1]$.}
\vskip 0.2cm

The next lemma is a version of the Wiener theorem (see, e.g., [45, Theorem 
XI.114 and the remark after it]).
\vskip 0.2cm

{\bf Lemma 6.1}. {\it Suppose $W\in L^1(K)$, and $\Psi $ is a real-valued function 
of class $\widetilde C(K)$ satisfying (5.1) for all $\lambda \in {\bf R}$. Then
$$
\lim\limits_{N\to +\infty }\, \frac 1N \sum\limits_{\nu =1}^N\, \biggl| \,
\int\limits_Ke^{\, 2\pi i\nu \, (\Psi - x_2)}\, W\, d^2x\, \biggr| ^2=0\, .
$$}
\vskip 0.2cm

{\bf Corollary}. {\it Under the conditions of Lemma 6.1, denote
$$
{\bf M}_{\pm}(W,\Psi ;\theta )=\biggl\{ \nu \in {\bf N}:\biggl| \, \int\limits
_Ke^{\, \pm 2\pi i\nu \, (\Psi -x_2)}\, W\, d^2x\biggr| \geq \theta \biggr\} 
\, ,\ \theta >0\, .
$$
Then ${\mathcal Q}\, ({\bf M}_{\pm}(W,\Psi ;\theta ))=0$ (for any $\theta >0$).}
\vskip 0.2cm

{\it Proof of Theorem 6.1}. Let $f_{V^{(\pm )}}$, $h_{V^{(\pm )}}$, and $c_7\, 
(V^{(\pm )})$ be as in Lemmas 4.2 and 4.3. We denote 
$$
c^{\, \prime}_7=\max \, \{ c_7(V^{\, (+)}),c_7(V^{\, (-)})\} \, ,\ c_8^{\, 
\prime}=\frac 16\, c_1^2(c_1+4(c_7^{\, \prime})^2)^{-1}
$$ 
(then $c_8^{\, \prime}\in (0,\frac 16\, c_1]$). Suppose $\mu \geq \mu _0>0$, where 
$\mu _0$ is a sufficiently large number to be chosen later. To start with, we
assume that $\mu _0\geq 4\pi $. Let $\varphi _{\pm}\in \widetilde H^1(K)$, and let
$k\in {\bf R}^2$. For $a\in [2\pi ,\mu /2]$, we denote
$$
\varphi ^{(a)}_{\pm}=\widehat P^{\, T^{\, \pm}(a)}\varphi _{\pm}\, ,\ \widetilde
\varphi ^{\, (a)}_{\pm}=\widehat P^{\, {\bf Z}^2\backslash T^{\, \pm}(a)}\varphi
_{\pm}\, ,
$$ 
where $T^{\, \pm}(a)=\{ N\in {\bf Z}^2: G^{\pm}_N(k;\mu )\leq a\} $ (the functions
$G^{\pm}_N(k;\mu )$ and $G_N(k;\mu )$, $N\in {\bf Z}^2$, and also the norms
$\| .\| _*$ and $\| .\| _{*,\pm}\, $, were defined at the beginning of {\S}{\,}4). 
Lemma 4.1 yields the estimates
$$
c_1\, \| \varphi ^{(a)}_{\pm}\| ^2_*=
c_1\, \| \varphi ^{(a)}_{\pm}\| ^2_{*,\pm}\leq \| (\widehat d_{\pm}(k)+i\mu
{\mathcal H})\varphi ^{(a)}_{\pm}\| ^2\leq
c_2\, \| \varphi ^{(a)}_{\pm}\| ^2_{*,\pm}=c_2\, \| \varphi ^{(a)}_{\pm}\| ^2_*\, ,
\eqno (6.1)
$$ $$
c_1\, \| \widetilde \varphi ^{\, (a)}_{\pm}\| ^2_{*,\pm}\leq \| (\widehat d_{\pm}(k)
+i\mu {\mathcal H})\widetilde \varphi ^{(a)}_{\pm}\| ^2\leq c_2\, \| \widetilde 
\varphi ^{\, (a)}_{\pm}\| ^2_{*,\pm}\, .  \eqno (6.2)
$$
We choose a number  $a_0^{\, \prime}\geq 2\pi $ for which
$$
\max\ \{ h^2_{V^{(+)}}(a_0^{\, \prime}),h^2_{V^{(-)}}(a_0^{\, \prime})\} 
\leq \frac 16\, c_8^{\, \prime}  \eqno (6.3)
$$
(this can be done because $h_{V^{(\pm)}}(t)\to 0$ as $t\to +\infty $). Let $a_1
\geq a_0^{\, \prime}\, $, and let $\delta = \min \, \{ \frac 1{32}\, 
c_1, \frac 32\, c_8^{\, \prime}\, \} $. We denote by $J$ the smallest integer with 
$c_2^2\leq J\delta ^2$ and choose numbers $a_2,\dots , a_{J+1}$ so that 
$a_{j+1}>a_j$ ($j=1,\dots ,J$) and each of the (four) functions ${\mathcal P}=
{\mathcal G}^2+{\mathcal F}^2$, ${\mathcal P}=({\mathcal G}\pm i{\mathcal F})
{\mathcal H}$, and ${\mathcal P}={\mathcal H}^2$ in the space $L^{\infty}(K)
\subset L^2(K)$ satisfies the inequalities
$$
a_j\, \biggl( \, \sum\limits_{N\in {\bf Z}^2\, :\, 2\pi |N|\, >\, a_{j+1}-a_j}|
{\mathcal P}_N|^2\biggr) ^{1/2}\leq \frac {\delta}{4\, \sqrt {6\pi }}\, , 
\eqno (6.4)
$$
where the ${\mathcal P}_N$ are the Fourier coefficients of ${\mathcal P}$
(the numbers $a_0^{\, \prime}$, $\delta$, and $J$ depend on $p$, $q$, $F$, $V^{(+)}$ 
and $V^{(-)}$, and the numbers $a_2,\dots ,a_{J+1}$ depend on $a_1$ and the
functions ${\mathcal F}$, ${\mathcal G}$, ${\mathcal H}$, $V^{(+)}$, and $V^{(-)}$). 
Now we specify the choice of $\mu _0\, $. We assume that $\mu _0\geq 2a_{J+1}$,
$$
\max\ \{ h^2_{V^{(+)}}(\mu _0),h^2_{V^{(-)}}(\mu _0)\} \leq \frac 1{54}\, 
c_8^{\, \prime}\, , \eqno (6.5)
$$ 
and, for all $\mu \geq \mu _0$ (and all $k\in {\bf R}^2$),
$$
\mu ^{-2}\, \max\  \{ f^2_{V^{(-)}}(\# \, T^{+}(\mu /2)),f^2_{V^{(+)}}
(\# \, T^{-}(\mu /2))\} \leq \frac 1{24}\, c_8^{\, \prime}  \eqno (6.6)
$$
(by Lemma 4.2 and the estimate $\# \, T^{\pm }(\mu /2)< \frac {3\pi }2
\, \mu ^2$ (see (4.1) with $a=\mu /2\geq \mu _0/2\geq 2\pi $), condition (6.6) 
can indeed be ensured if $\mu _0$ is chosen sufficiently large). Since
$$
\sum\limits_{j=1}^J\| \widetilde \varphi _{\pm}^{(a_j)}-\widetilde \varphi 
_{\pm}^{(a_{j+1})}\| ^2_{*,\pm}=\| \widetilde \varphi _{\pm}^{(a_1)}-\widetilde 
\varphi _{\pm}^{(a_{J+1})}\| ^2_{*,\pm}=\| \widetilde \varphi _{\pm}^{(a_1)}-\widetilde 
\varphi _{\pm}^{(a_{J+1})}\| ^2_*\leq \| \varphi _{\pm}\| ^2_*\, ,
$$
it follows that, depending on the functions $\varphi _{\pm}\in \widetilde H^1(K)$,
we can find indices $j_{\pm}\in \{ 1,\dots ,J\} $ such that
$$
\| \widetilde \varphi _{\pm}^{(a_{j_{\pm}})}-\widetilde \varphi _{\pm}^{(a_{j_{\pm}+1})}\| 
^2_{*,\pm}\leq J^{-1}\, \| \varphi _{\pm}\| ^2_*\, .
$$
Then, by Lemma 4.1 and (6.2), we have
$$
\| (\widehat d_{\pm}(k)+i\mu {\mathcal H})(\widetilde \varphi _{\pm}^{(a_{j_{\pm}})}
-\widetilde \varphi _{\pm}^{(a_{j_{\pm}+1})})\| \leq  \eqno (6.7)
$$ $$
\leq \sqrt {c_2}\, \| \widetilde \varphi _{\pm}^{(a_{j_{\pm}})}-\widetilde 
\varphi _{\pm}^{(a_{j_{\pm}+1})}\| _{*,\pm}\leq \biggl( \frac {c_2}J\biggr) ^{1/2}
\| \varphi _{\pm}\| _*\, .
$$
The above choice of the indices $j_{\pm}$ is important for the proof of the next
lemma.
\vskip 0.2cm

{\bf Lemma 6.2}. {\it We have
$$
|((\widehat d_{\pm}(k)+i\mu {\mathcal H})\widetilde \varphi _{\pm}^{(a_{j_{\pm}})},
(\widehat d_{\pm}(k)+i\mu {\mathcal H})\varphi _{\pm}^{(a_{j_{\pm}})})|\leq 
2\delta\, \| \varphi _{\pm}\| _{*,\pm}\, \| \varphi _{\pm} ^{(a_{j_{\pm}})}\| _*\ .  
\eqno (6.8)
$$}

{\it Proof}. From (6.7) we deduce that
$$
|((\widehat d_{\pm}(k)+i\mu {\mathcal H})\widetilde \varphi _{\pm}^{(a_{j_{\pm}})},
(\widehat d_{\pm}(k)+i\mu {\mathcal H})\varphi _{\pm}^{(a_{j_{\pm}})})|\leq  
\eqno (6.9)
$$ $$
\leq |((\widehat d_{\pm}(k)+i\mu {\mathcal H})\widetilde 
\varphi _{\pm}^{(a_{j_{\pm}+1})}, (\widehat d_{\pm}(k)+i\mu {\mathcal H})
\varphi _{\pm}^{(a_{j_{\pm}})})|+  
$$ $$  
+|((\widehat d_{\pm}(k)+i\mu {\mathcal H})(\widetilde \varphi _{\pm}^{(a_{j_{\pm}})}
-\widetilde \varphi _{\pm}^{(a_{j_{\pm}+1})}), (\widehat d_{\pm}(k)+i\mu {\mathcal H})
\varphi _{\pm}^{(a_{j_{\pm}})})|\leq  
$$ $$
\leq |((\widehat d_{\pm}(k)+i\mu {\mathcal H})\widetilde 
\varphi _{\pm}^{(a_{j_{\pm}+1})}, (\widehat d_{\pm}(k)+i\mu {\mathcal H})
\varphi _{\pm}^{(a_{j_{\pm}})})|+
$$ $$
+\biggl( \frac {c_2}J\biggr) ^{1/2}\| \varphi _{\pm}
\| _*\, \| (\widehat d_{\pm}(k)+i\mu {\mathcal H})\varphi _{\pm}^{(a_{j_{\pm}})}\| \, .
$$
Also, we have $\| \varphi _{\pm}\| _*\leq \| \varphi _{\pm}\| _{*,\pm}$ and
$$
\| (\widehat d_{\pm}(k)+i\mu {\mathcal H})\varphi ^{(a_{j_{\pm}})}_{\pm}\| \leq
\sqrt c_2\, \| \varphi ^{(a_{j_{\pm}})}_{\pm}\| _{*,\pm}=\sqrt c_2\, 
\| \varphi ^{(a_{j_{\pm}})}_{\pm}\| _*
$$
by (6.1). Since
$$
\widehat d_{\pm}(k)+i\mu {\mathcal H}=
({\mathcal G}\pm i{\mathcal F})\bigl( k_1-i\frac {\partial}{\partial x_1}\bigr) 
+i{\mathcal H}\bigl( \mu \pm \bigl( k_2-i\frac {\partial}{\partial x_2}\bigr) 
\bigr)\, ,
$$
we can write (see Lemma 4.5 and inequalities (6.4))
$$
|((\widehat d_{\pm}(k)+i\mu {\mathcal H})\widetilde 
\varphi _{\pm}^{(a_{j_{\pm}+1})}, (\widehat d_{\pm}(k)+i\mu {\mathcal H})
\varphi _{\pm}^{(a_{j_{\pm}})})|\leq
$$ $$
\leq \sqrt {6\pi }\, a_{j_{\pm}}\, \biggl( \, \sum\limits_{N\in {\bf Z}^2\, :\, 2\pi |N|\, >\,
a_{j_{\pm}+1}-a_{j_{\pm}}}|({\mathcal G}^2+{\mathcal F}^2)_N|^2\biggr) ^{1/2}
\times
$$ $$
\times \ \bigl\| \bigl( k_1-i\frac {\partial}{\partial x_1}\bigr) \widetilde
\varphi _{\pm}^{\, (a_{j_{\pm}+1})}\bigr\| \cdot \bigl\| \bigl( k_1-i\frac
{\partial}{\partial x_1}\bigr) \varphi _{\pm} ^{(a_{j_{\pm}})}\bigr\| +
$$ $$
+\sqrt {6\pi }\, a_{j_{\pm}}\, \biggl( \, \sum\limits_{N\in {\bf Z}^2\, :\, 2\pi |N|\, >\,
a_{j_{\pm}+1}-a_{j_{\pm}}}|(({\mathcal G}\mp i{\mathcal F}){\mathcal H})_N|^2
\biggr) ^{1/2}\times
$$ $$
\times \ \bigl\| \bigl( k_1-i\frac {\partial}{\partial x_1}\bigr) \widetilde
\varphi _{\pm}^{\, (a_{j_{\pm}+1})}\bigr\| \cdot \bigl\| \bigl( \mu \pm \bigl( 
k_2-i\frac {\partial}{\partial x_2}\bigr) \bigr) \varphi _{\pm} ^{(a_{j_{\pm}})}
\bigr\| +
$$ $$
+\sqrt {6\pi }\, a_{j_{\pm}}\, \biggl( \, \sum\limits_{N\in {\bf Z}^2\, :\, 2\pi |N|\, >\,
a_{j_{\pm}+1}-a_{j_{\pm}}}|(({\mathcal G}\pm i{\mathcal F}){\mathcal H})_N|^2
\biggr) ^{1/2}\times
$$ $$
\times \ \bigl\| \bigl( \mu \pm \bigl( k_2-i\frac {\partial}{\partial x_2}
\bigr) \bigr) \widetilde \varphi _{\pm}^{\, (a_{j_{\pm}+1})}\bigr\| \cdot \bigl\| 
\bigl( k_1-i\frac {\partial}{\partial x_1}\bigr) \varphi _{\pm} ^{(a_{j_{\pm}})}
\bigr\| +
$$ $$
+\sqrt {6\pi }\, a_{j_{\pm}}\, \biggl( \, \sum\limits_{N\in {\bf Z}^2\, :\, 2\pi |N|\, >\,
a_{j_{\pm}+1}-a_{j_{\pm}}}|({\mathcal H}^2)_N|^2\biggr) ^{1/2}\times
$$ $$
\times \ \bigl\| \bigl( \mu \pm \bigl( k_2-i\frac {\partial}{\partial x_2}
\bigr) \bigr) \widetilde \varphi _{\pm}^{\, (a_{j_{\pm}+1})}\bigr\| \cdot 
\bigl\| \bigl( \mu \pm \bigl( k_2-i\frac {\partial}{\partial x_2}\bigr) 
\bigr) \varphi _{\pm} ^{(a_{j_{\pm}})}\bigr\| \leq
$$ $$
\leq \delta \, \| \widetilde \varphi _{\pm}^{\, (a_{j_{\pm}+1})}\| _{*,\pm}\,
\| \varphi _{\pm} ^{(a_{j_{\pm}})}\| _{*,\pm}\leq \delta \, \| \varphi _{\pm}
\| _{*,\pm}\, \| \varphi _{\pm} ^{(a_{j_{\pm}})}\| _*\, .
$$
Now, estimate (6.8) for all $\mu \geq \mu _0$ and all $k\in {\bf R}^2$ follows
from (6.9) and the choice of $J\in {\bf N}$. Lemma 6.2 is proved.
\hfill  $\square$
\vskip 0.2cm

Now we define a set ${\bf M}\subset {\bf N}$ (for which ${\mathcal Q}\, ({\bf N}
\backslash {\bf M})=0$). We have $\# \, \{ N\in {\bf Z}^2:\pi \, |N|<a_J\} \leq 
{\tau}^*\doteq 4{\pi }^{-1}a_J^2\, $. Put $\theta =\frac 13\, 2^{-6}c_1(\pi 
a_1^2{\tau}^*)^{-1}$. The corollary to Lemma 6.1 implies that for all functions 
${\mathcal P}^{\, \prime}_{\pm}=({\mathcal G}\pm i{\mathcal F})V^{\, (\pm )}e^{\, 
2\pi i\, (N,\, x)}$ and ${\mathcal P}^{\, \prime}_{\pm}={\mathcal H}V^{\, (\pm )}
e^{\, 2\pi i\, (N,\, x)}$, where $N\in {\bf Z}^2$ and $\pi \, |N|<a_J\, $, we have 
${\mathcal Q}\, ({\bf M}_{\pm}({\mathcal P}^{\, \prime}_{\pm}, \Psi ; \theta ))=0$. 
Let ${\bf M}^{\, \prime}$ be the union of the sets ${\bf M}_{\pm}({\mathcal P}^{\, 
\prime}_{\pm},\Psi ; \theta )$ for both signs $+$ and $-$ and for all functions 
${\mathcal P}^{\, \prime}_{\pm}\, $ as above. Since there are at most $4{\tau}^*$
such sets, we have ${\mathcal Q}\, ({\bf M}^{\, \prime})=0$. We put ${\bf M}={\bf N}
\backslash ({\bf M}^{\, \prime}\cup \{ m\in {\bf N}: \pi m <\mu _0\} )$. Then 
${\mathcal Q}\, ({\bf N}\backslash {\bf M})=0$, and $\mu \geq \mu _0$ whenever 
$\mu \in \pi {\bf M}$. In what follows, we assume that $\mu \in \pi {\bf M}$.
\vskip 0.2cm

{\bf Lemma 6.3}. {\it For all $\mu \in \pi {\bf M}$ (and all $k\in {\bf R}^2$)
we have
$$
|((\widehat d_{\pm}(k)+i\mu {\mathcal H})\varphi _{\pm}^{(a_{j_{\pm}})}, e^{\, 
\mp 2i\mu \Psi }\, V^{\, (\mp )}\varphi _{\mp}^{(a_1)})|\leq \frac {c_1}{16}\, 
\| \varphi _{\pm}^{(a_{j_{\pm}})}\| _*\, \| \varphi _{\mp}^{(a_1)}\| \, . \eqno (6.10)
$$}

{\it Proof}. If $\mu \in \pi {\bf M}$, then ${\mu}/{\pi}\in
{\bf N}\backslash {\bf M}_{\pm}({\mathcal P}^{\, \prime}_{\pm},\Psi ; \theta )$
for all functions ${\mathcal P}^{\, \prime}_{\pm}\, $ indicated above.
Therefore, the definition of the set ${\bf M}_{\pm}({\mathcal P}^{\, \prime}_{\pm},
\Psi ; \theta )$ implies the inequalities  
$$
6\pi a_1^2\sum\limits_{N\in {\bf Z}^2\, :\, \pi |N|\, <\, a_J}\, \biggl| \, 
\int\limits_Ke^{\, \mp 2i\mu \, (\Psi -x_2)}({\mathcal G}\mp i{\mathcal F})V^{\, 
(\mp )}e^{\, 2\pi i\, (N,\, x)}d^2x\, \biggr| <
$$ $$
<6\pi a_1^2{\tau}^*\theta =\frac {c_1}{32}
\, ,
$$ $$
6\pi a_1^2\sum\limits_{N\in {\bf Z}^2\, :\, \pi |N|\, <\, a_J}\, \biggl| \, 
\int\limits_Ke^{\, \mp 2i\mu \, (\Psi -x_2)}\, {\mathcal H}\, V^{\, (\mp )}
e^{\, 2\pi i\, (N,\, x)}d^2x\, \biggr| <
$$ $$
<6\pi a_1^2{\tau}^*\theta =\frac {c_1}{32}\, .
$$
Since $\# \, T^{\pm }(a_1)\leq 6\pi a_1^2$, we deduce the estimate
$$
|((\widehat d_{\pm}(k)+i\mu {\mathcal H})\varphi _{\pm}^{(a_{j_{\pm}})}, e^{\, 
\mp 2i\mu \Psi }\, V^{\, (\mp )}\varphi _{\mp}^{(a_1)})|\leq  
$$ $$
\leq \bigl| \bigl( ({\mathcal G}\pm i{\mathcal F})\bigl( k_1-i\frac {\partial}
{\partial x_1}\bigr) \varphi _{\pm}^{(a_{j_{\pm}})}, e^{\, \mp 2i\mu \Psi }\, V^{\, 
(\mp )}\varphi _{\mp}^{(a_1)}\bigr) \bigr| +
$$ $$
+\bigl| \bigl( {\mathcal H}\bigl( \mu \pm \bigl( k_2-i\frac {\partial}{\partial 
x_2}\bigr) \bigr) \varphi _{\pm}^{(a_{j_{\pm}})}, e^{\, \mp 2i\mu \Psi }\, V^{\, 
(\mp )}\varphi _{\mp}^{(a_1)}\bigr) \bigr| \leq
$$ $$
\leq \biggl| \ \sum\limits_{N\in \, T^{\, \pm}(a_{j_{\pm}}),\, M\in \, T^{\,
\mp}(a_1)} \ (k_1+2\pi N_1)\, \overline {(\varphi
_{\pm})_N}\, (\varphi _{\mp})_M \times
$$ $$
\times \int\limits_K e^{\, \mp 2i \mu \Psi +2\pi
i\, (M-N,\, x)}({\mathcal G}\mp i{\mathcal F})V^{\, (\mp )}d^2x\ \biggr| \, +
$$ $$
+\, \biggl| \sum\limits_{N\in \, T^{\, \pm}(a_{j_{\pm}}),\, M\in \, T^{\, \mp}
(a_1)}(\mu \pm (k_2+2\pi N_2))\overline {(\varphi _{\pm})_N}(\varphi
_{\mp})_M \times
$$ $$
\times \int\limits_K e^{\, \mp 2i \mu \Psi +2\pi i\, (M-N,\, x)}{\mathcal H}
V^{\, (\mp )}d^2x\, \biggr| \leq
$$ $$
\leq 6\pi a_1^2\, \| \varphi _{\pm}^{(a_{j_{\pm}})}\| _*\, \| \varphi _{\mp}^{(a_1)}
\| \times
$$ $$
\times \biggl( \, \sum\limits_{N^{\, \prime}\in {\bf Z}^2\, :\, \pi |N^{\,
\prime}| \, <\, a_J} \biggl| \, \int\limits^{\phantom a}_K e^{\, \mp 2i\mu \, 
(\Psi -x_2)}({\mathcal G}\mp i{\mathcal F})V^{\, (\mp )}e^{\, 2\pi i\, (N^{\,
\prime},\, x)}d^2x\, \biggr| +
$$ $$
+\sum\limits_{N^{\, \prime}\in {\bf Z}^2\, :\, \pi |N^{\, \prime}|
\, <\, a_J} \biggl| \, \int\limits_K e^{\, \mp 2i\mu \, (\Psi -x_2)}\, {\mathcal H}
\, V^{\, (\mp )}e^{\, 2\pi i\, (N^{\, \prime},\, x)}d^2x\, \biggr| \, \biggr) \leq
$$ $$
\leq \frac {c_1}{16}\, \| \varphi _{\pm}^{(a_{j_{\pm}})}\| _*\, \| \varphi
_{\mp}^{(a_1)}\| \, . \phantom {\int}
$$
This proves Lemma 6.3.   \hfill  $\square$
\vskip 0.2cm

We pass to the proof of the main estimates. In the sequel it is assumed that 
$k\in {\bf R}^2$ is such that $k_1=\pi $. Taking into account the inequalities 
$$
\| \varphi _{\mp}^{(a_1)}\| \leq \| \varphi _{\mp}^{(a_{j_{\mp}})}\| \leq
\pi ^{-1}\, \| \varphi _{\mp}^{(a_{j_{\mp}})}\| _*\leq \| \varphi _{\mp}^{(a_{j
_{\mp}})}\| _*\, ,
$$
and using (6.8) and (6.10), we obtain
$$
\| (\widehat d_{\pm}(k)+i\mu {\mathcal H})\varphi _{\pm}+e^{\, 
\mp 2i\mu \Psi }\, V^{\, (\mp )}\varphi _{\mp}^{(a_1)}\| ^2=
$$ $$
=\| (\widehat d_{\pm}(k)+i\mu {\mathcal H})\varphi _{\pm}^{(a_{j_{\pm}})}
+(\widehat d_{\pm}(k)+i\mu {\mathcal H})\widetilde \varphi _{\pm}^{\, (a_{j_{\pm}})}
+e^{\, \mp 2i\mu \Psi }\, V^{\, (\mp )}\varphi _{\mp}^{(a_1)}\| ^2\geq
$$ $$
\geq \| (\widehat d_{\pm}(k)+i\mu {\mathcal H})\varphi _{\pm}^{(a_{j_{\pm}})}\| ^2
+\| (\widehat d_{\pm}(k)+i\mu {\mathcal H})\widetilde \varphi _{\pm}^{\, (a_{j_{\pm}})}
+e^{\, \mp 2i\mu \Psi }\, V^{\, (\mp )}\varphi _{\mp}^{(a_1)}\| ^2-
$$ $$
-2\, |((\widehat d_{\pm}(k)+i\mu {\mathcal H})\widetilde 
\varphi _{\pm}^{\, (a_{j_{\pm}})}, (\widehat d_{\pm}(k)+i\mu {\mathcal H})
\varphi _{\pm}^{(a_{j_{\pm}})})|-
$$ $$
-2|((\widehat d_{\pm}(k)+i\mu {\mathcal H})\varphi _{\pm}^{(a_{j_{\pm}})}, e^{\, 
\mp 2i\mu \Psi }\, V^{\, (\mp )}\varphi _{\mp}^{(a_1)})|\geq
$$ $$
\geq c_1\, \| \varphi _{\pm}^{(a_{j_{\pm}})}\| ^2_*+\| (\widehat d_{\pm}(k)+i\mu 
{\mathcal H})\widetilde \varphi _{\pm}^{\, (a_{j_{\pm}})}+e^{\, \mp 2i\mu \Psi }\, 
V^{\, (\mp )}\varphi _{\mp}^{(a_1)}\| ^2-
$$ $$
-4\delta \, \| \varphi _{\pm}\| _{*,\pm}\, \| \varphi _{\pm}^{(a_{j_{\pm}})}\| _*
-\frac {c_1}8\, \| \varphi _{\pm}^{(a_{j_{\pm}})}\| _*\, \| \varphi _{\mp}^{(a_1)}\|
\geq
$$ $$
\geq c_1\, \| \varphi _{\pm}^{(a_{j_{\pm}})}\| ^2_*+\| (\widehat d_{\pm}(k)+i\mu 
{\mathcal H})\widetilde \varphi _{\pm}^{\, (a_{j_{\pm}})}+e^{\, \mp 2i\mu \Psi }\, 
V^{\, (\mp )}\varphi _{\mp}^{(a_1)}\| ^2-
$$ $$ 
-2\delta \, \bigl( \, \| \widetilde \varphi _{\pm}^{\, (a_{j_{\pm}})}\|
^2_{*,\pm}+ 2\, \| \varphi _{\pm}^{(a_{j_{\pm}})}\| ^2_{*,\pm}\, \bigr) -\frac
{c_1}{16}\, \bigl( \, \| \varphi _{\pm}^{(a_{j_{\pm}})}\| ^2_*+\| \varphi
_{\mp}^{(a_{j_{\mp}})}\| ^2_* \, \bigr) \, .
$$
For any $\varepsilon \in (0,1)$, we have (see Lemma 4.3)
$$
\| (\widehat d_{\pm}(k)+i\mu {\mathcal H})\widetilde 
\varphi _{\pm}^{\, (a_{j_{\pm}})}+e^{\, \mp 2i\mu \Psi }\, 
V^{\, (\mp )}\varphi _{\mp}^{(a_1)}\| ^2\geq
$$ $$
\geq (1-\varepsilon )\, \| (\widehat d_{\pm}(k)+i\mu 
{\mathcal H})\widetilde \varphi _{\pm}^{\, (a_{j_{\pm}})}\| ^2-
(1-\varepsilon )\varepsilon ^{-1}\| V^{\, (\mp )}\varphi _{\mp}^{(a_1)}\| ^2\geq
$$ $$
\geq (1-\varepsilon )c_1\, \| \widetilde \varphi _{\pm}^{\, (a_{j_{\pm}})}\|
^2_{*,\pm}- (1-\varepsilon )\varepsilon ^{-1}(c_7(V^{(\mp )}))^2\, \| \varphi
_{\mp}^{(a_1)}\| ^2_*\, \geq
$$ $$
\geq (1-\varepsilon )c_1\, \| \widetilde \varphi _{\pm}^{\, (a_{j_{\pm}})}\|
^2_{*,\pm}- (1-\varepsilon )\varepsilon ^{-1}(c_7^{\, \prime})^2\, \| \varphi
_{\mp}^{(a_{j_{\mp}})}\| ^2_*\, .
$$
Therefore,
$$
\sum\limits_{+,-}\| (\widehat d_{\pm}(k)+i\mu {\mathcal H})\varphi _{\pm}
+e^{\, \mp 2i\mu \Psi }\, V^{\, (\mp )}\varphi _{\mp}^{(a_1)}\| ^2\geq 
$$ $$
\geq \bigl( c_1-(1-\varepsilon )\varepsilon ^{-1}(c_7^{\, \prime})^2-4\delta 
-\frac {c_1}8\, \bigr) \sum\limits_{+,-}\| \varphi _{\pm}^{(a_{j_{\pm}})}\| ^2_* +
\bigl( (1-\varepsilon )c_1-2\delta \bigr) \sum\limits_{+,-} \| \widetilde \varphi
_{\pm}^{\, (a_{j_{\pm}})}\| ^2_{*,\pm}\, .
$$
If $c_7^{\, \prime}>0$, we put $\varepsilon =4\, (c_7^{\, \prime})^2(c_1
+4\, (c_7^{\, \prime})^2)^{-1}$. Since $(1-\varepsilon )\varepsilon ^{-1} (c_7^{\, 
\prime})^2=c_1/4$, $4\delta \leq c_1/8$ and $2\delta \leq \frac 12\, (1-\varepsilon )
c_1=3c_8^{\, \prime}\, $, we have
$$
\sum\limits_{+,-}\| (\widehat d_{\pm}(k)+i\mu {\mathcal H})\varphi _{\pm}
+e^{\, \mp 2i\mu \Psi }\, V^{\, (\mp )}\varphi _{\mp}^{(a_1)}\| ^2\geq \eqno (6.11)
$$ $$
\geq \frac {c_1}2\,
\sum\limits_{+,-}\| \varphi _{\pm}^{(a_{j_{\pm}})}\| ^2_*+3c_8^{\, \prime}
\, \sum\limits_{+,-} \| \widetilde \varphi
_{\pm}^{\, (a_{j_{\pm}})}\| ^2_{*,\pm}\, , 
$$
which is also true if $c_7^{\, \prime}=0$. Estimate (6.11) is a key point in the
proof of Theorem 6.1, and now it is not too hard to complete the proof. For 
$\varphi _{\pm}\in \widetilde H^1(K)$, we have
$$
\widetilde \varphi ^{\, (a_1)}_{\pm}=\widehat P^{\, T^{\pm}(\mu /2)\backslash
T^{\pm}(a_1)}\varphi _{\pm}+\widehat P^{\, {\bf Z}^2\backslash (T^+(\mu /2)\cup
T^-(\mu /2))}\varphi _{\pm}+\widehat P^{\, T^{\mp}(\mu /2)}\varphi _{\pm}\, .
$$
Using Lemmas 4.2, 4.3, and 4.4, and conditions (6.3), (6.5), and (6.6), we obtain
$$
\| V^{\, (\mp )}\widehat P^{\, T^{\mp}(\mu /2)\backslash
T^{\mp}(a_1)}\varphi _{\mp}\| ^2\leq h^2_{V^{(\mp )}}(a_1) \, \| \widehat 
P^{\, T^{\mp}(\mu /2)\backslash T^{\mp}(a_1)}\varphi _{\mp}\| ^2_*\leq
$$ $$
\leq \frac 16\, c_8^{\, \prime}\, \| \widehat 
P^{\, T^{\mp}(\mu /2)\backslash T^{\mp}(a_1)}\varphi _{\mp}\| ^2_*\, ,
$$ $$
\| V^{\, (\mp )}\widehat P^{\, {\bf Z}^2\backslash (T^+(\mu /2)\cup
T^-(\mu /2))}\varphi _{\mp}\| ^2\leq 9h^2_{V^{(\mp )}}(\mu ) \, \| \widehat P^{\, 
{\bf Z}^2\backslash (T^+(\mu /2)\cup T^-(\mu /2))}\varphi _{\mp}\| ^2_*\leq
$$ $$
\leq \frac 16\, c_8^{\, \prime}\, \| \widehat P^{\, 
{\bf Z}^2\backslash (T^+(\mu /2)\cup T^-(\mu /2))}\varphi _{\mp}\| ^2_*\, ,
$$ $$
\| V^{\, (\mp )}\widehat P^{\, T^{\pm}(\mu /2)}\varphi _{\mp}\| ^2\leq 
f^2_{V^{(\mp )}}(\# \, T^{\pm }(\mu /2)) \, \| \widehat P^{\, T^{\pm}(\mu /2)}
\varphi _{\mp}\| ^2\leq
$$ $$
\leq 4\mu ^{-2}\, f^2_{V^{(\mp )}}(\# \, T^{\pm }(\mu /2)) \, \| \widehat P^{\, 
T^{\pm}(\mu /2)}\varphi _{\mp}\| ^2_{*,\mp}\leq \frac 16\, c_8^{\, \prime}\,
\| \widehat P^{\, T^{\pm}(\mu /2)}\varphi _{\mp}\| ^2_{*,\mp}\, .
$$
Therefore,
$$
\sum\limits_{+,-}\| (\widehat d_{\pm}(k)+i\mu {\mathcal H})\varphi _{\pm}
+e^{\, \mp 2i\mu \Psi }\, V^{\, (\mp )}\varphi _{\mp}\| ^2\geq  \eqno (6.12)
$$ $$
\geq \frac 12\, \sum\limits_{+,-}\| (\widehat d_{\pm}(k)+i\mu {\mathcal H})
\varphi _{\pm}+e^{\, \mp 2i\mu \Psi }\, V^{\, (\mp )}\varphi _{\mp}^{(a_1)}\| ^2-  
$$ $$
-3\, \sum\limits_{+,-}\| V^{\, (\mp )}\widehat P^{\, T^{\mp}(\mu /2)\backslash
T^{\mp}(a_1)}\varphi _{\mp}\| ^2-
3\, \sum\limits_{+,-}\| V^{\, (\mp )}\widehat P^{\, {\bf Z}^2\backslash 
(T^+(\mu /2)\cup T^-(\mu /2))}\varphi _{\mp}\| ^2-
$$ $$
-3\, \sum\limits_{+,-}\| V^{\, (\mp )}\widehat P^{\, T^{\pm}(\mu /2)}
\varphi _{\mp}\| ^2\geq
$$ $$
\geq \frac 12\, \sum\limits_{+,-}\| (\widehat d_{\pm}(k)+i\mu {\mathcal H})
\varphi _{\pm}+e^{\, \mp 2i\mu \Psi }\, V^{\, (\mp )}\varphi _{\mp}^{(a_1)}\| ^2-
\frac 12\, c_8^{\, \prime}\, \sum\limits_{+,-}\|\widetilde \varphi _{\mp}^{\,
(a_1)}\| ^2_{*,\mp}\, .
$$
Finally, from (6.11), (6.12), and the condition $c_8^{\, \prime}\leq \frac 16\, 
c_1$ we deduce the estimate claimed in Theorem 6.1:
$$
\sum\limits_{+,-}\| (\widehat d_{\pm}(k)+i\mu {\mathcal H})\varphi _{\pm}
+e^{\, \mp 2i\mu \Psi }\, V^{\, (\mp )}\varphi _{\mp}\| ^2\geq 
$$ $$
\geq \frac {c_1}4\, \sum\limits_{+,-}\| \varphi _{\pm}^{(a_{j_{\pm}})}\| ^2_{*,\pm}
+\frac 32\, c_8^{\, \prime}\, \sum\limits_{+,-} \| \widetilde \varphi _{\pm}^{\, 
(a_{j_{\pm}})}\| ^2_{*,\pm}-\frac 12\, c_8^{\, \prime}\, \sum\limits_{+,-}
\| \widetilde \varphi _{\mp}^{\, (a_1)}\| ^2_{*,\mp}\geq
$$ $$
\geq \frac {c_1}6\, 
\sum\limits_{+,-}\| \varphi _{\pm}^{(a_{j_{\pm}})}\| ^2_{*,\pm}+
c_8^{\, \prime}\, \sum\limits_{+,-} \| \widetilde \varphi
_{\pm}^{\, (a_{j_{\pm}})}\| ^2_{*,\pm}+\frac 12\, c_8^{\, \prime}\,
\sum\limits_{+,-} \| \varphi _{\pm}^{\, (a_1)}\| ^2_{*,\pm}\geq
$$ $$
\geq \frac {c_1}6\, 
\sum\limits_{+,-}\| \varphi _{\pm}^{(a_{j_{\pm}})}\| ^2_*+
c_8^{\, \prime}\, \sum\limits_{+,-} \| \widetilde \varphi
_{\pm}^{\, (a_{j_{\pm}})}\| ^2_{*,\pm} \geq
\frac {c_1}6\, 
\sum\limits_{+,-}\| \varphi _{\pm}^{(a_1)}\| ^2_*+
c_8^{\, \prime}\, \sum\limits_{+,-} \| \widetilde \varphi
_{\pm}^{\, (a_1)}\| ^2_{*,\pm}\, .
$$
Theorem 6.1 is proved.   \hfill $\square$


\begin{thebibliography}{99}

\bibitem{1} P. Kuchment, {\it Floquet theory for partial differential equations},
Oper. Theory Adv. Appl., vol.~60, Birkh\"{a}user Verlag, Basel, 1993. 

\bibitem{2} P. Kuchment and S. Levendorski$\breve {\rm i}$, {\it On the structure
of spectra of periodic elliptic operators}, Trans. Amer. Math. Soc. {\bf 354} (2002), 
no. 2, 537-569.

\bibitem{3} L. I. Danilov, {\it On the spectrum of the Dirac operator with
periodic potential}, Preprint, Fiz.-Tekhn. Inst. Ural. Otdel. Akad. Nauk SSSR, 
Sverdlovsk, 1987 (in Russian).

\bibitem{4} L. I. Danilov, {\it A property of the integer lattice in ${\bf R}^3$
and the spectrum of the Dirac operator with periodic potential}, Preprint, 
Fiz.-Tekhn. Inst. Ural. Otdel. Akad. Nauk SSSR, Sverdlovsk, 1988 (in Russian).

\bibitem{5} L. I. Danilov, {\it On the spectrum of the Dirac operator in ${\bf R}^n$ 
with periodic potential}, Teoret. Mat. Fiz. {\bf 85} (1990), no. 1, 41-53; English
transl., Theoret. and Math. Phys. {\bf 85} (1990), no. 1, 1039-1048.

\bibitem{6} L. I. Danilov, {\it The spectrum of the Dirac operator with
periodic potential. I}, Fiz.-Tekhn. Inst. Ural. Otdel. Akad. Nauk SSSR, Izhevsk, 1991.
Manuscript dep. VINITI 12.12.91, no. 4588-B91 (in Russian).

\bibitem{7} L. I. Danilov, {\it The spectrum of the Dirac operator with
periodic potential. III}, Fiz.-Tekhn. Inst. Ural. Otdel. Ross. Akad. Nauk, Izhevsk, 
1992. Manuscript dep. VINITI 10.07.92, no. 2252-B92 (in Russian).

\bibitem{8} L. I. Danilov, {\it The spectrum of the Dirac operator with
periodic potential. VI}, Fiz.-Tekhn. Inst. Ural. Otdel. Ross. Akad. Nauk, Izhevsk, 
1996. Manuscript dep. VINITI 31.12.96, no. 3855-B96 (in Russian).  

\bibitem{9} L. I. Danilov, {\it Resolvent estimates and the spectrum of the Dirac
operator with a periodic potential}, Teoret. Mat. Fiz. {\bf 103} (1995), no. 1, 3-22; 
English transl., Theoret. and Math. Phys. {\bf 103} (1995), no. 1, 349-365. 

\bibitem{10} L. I. Danilov, {\it Absolute continuity of the spectrum of a 
periodic Dirac operator}, Differentsial'nye Uravneniya {\bf 36} (2000), no. 2, 233-240; 
English transl., Differential Equations {\bf 36} (2000), no. 2, 262-271. 

\bibitem{11} L. I. Danilov, {\it On the spectrum of the two-dimensional periodic
Dirac operator}, Teoret. Mat. Fiz. {\bf 118} (1999), no. 1, 3-14; English
transl., Theoret. and Math. Phys. {\bf 118} (1999), no. 1, 1-11.

\bibitem{12} M. Sh. Birman and T. A. Suslina, {\it The periodic Dirac operator is
absolutely continuous}, Integral Equations and Operator Theory {\bf 34} (1999),
377-395. 

\bibitem{13} L. I. Danilov, {\it On the spectrum of the periodic
Dirac operator}, Teoret. Mat. Fiz. {\bf 124} (2000), no. 1, 3-17; English
transl., Theoret. and Math. Phys. {\bf 124} (2000), no. 1, 859-871.

\bibitem{14} M. Sh. Birman and T. A. Suslina, {\it Two-dimensional periodic magnetic
Hamiltonian is absolutely continuous}, Algebra i Analiz {\bf 9} (1997), no. 1, 
32-48; English transl., St. Petersburg Math. J. {\bf 9} (1998), no. 1, 21-32.

\bibitem{15} M. Sh. Birman and T. A. Suslina, {\it Absolute continuity of the
two-dimensional periodic magnetic Hamiltonian with discontinuous vector-valued
potential}, Algebra i Analiz {\bf 10} (1998), no. 4, 1-36; English transl., St. 
Petersburg Math. J. {\bf 10} (1999), no. 4, 579-601.

\bibitem{16} I. S. Lapin, {\it Absolute continuity of the spectra of two-dimensional
periodic magnetic Schr\" odinger operator and Dirac operator with potentials in
the Zygmund class}, Probl. Mat. Anal., vyp. 22, S.-Peterburg. Univ., St. Petersburg, 
2001, pp. 74-105; English transl., J. Math. Sci. (New York) {\bf 106} (2001), no. 3, 
2952-2974.

\bibitem{17} A. V. Sobolev, {\it Absolute continuity of the periodic magnetic
Schr\"{o}dinger operator}, Invent. Math. {\bf 137} (1999), 85-112. 

\bibitem{18} M. Sh. Birman and T. A. Suslina, {\it Periodic magnetic Hamiltonian
with variable metric. The problem of absolute continuity}, Algebra i Analiz {\bf 11} 
(1999), no. 2, 1-40; English transl., St. Petersburg Math. J. {\bf 11} (2000), no. 2, 
203-232. 

\bibitem{19} L. I. Danilov, {\it On absolute continuity of the spectrum of
periodic Schr\" odinger and Dirac operators. I}, Fiz.-Tekhn. Inst. Ural. Otdel. Ross. 
Akad. Nauk, Izhevsk, 2000. Manuscript dep. VINITI 15.06.00, no. 1683-B00 (in Russian).  

\bibitem{20} L. I. Danilov, {\it On absolute continuity of the spectrum of a
periodic Schr\" odinger operator}, Mat. Zametki {\bf 73} (2003), no. 1, 49-62; English
transl., Math. Notes {\bf 73} (2003), no. 1-2, 46-57.

\bibitem{21} Z. Shen, {\it On absolute continuity of the periodic
Schr\"{o}dinger operators}, Internat. Math. Res. Notices {\bf 2001}, no. 1, 1-31.

\bibitem{22} Z. Shen, {\it Absolute continuity of periodic Schr\"{o}dinger 
operators with potentials in the Kato class}, Illinois J. Math. {\bf 45} (2001),
no. 3, 873-893.

\bibitem{23} Z. Shen, {\it The periodic Schr\"{o}dinger operators with
potentials in the Morrey class}, J. Funct. Anal. {\bf 193} (2002), no. 2, 
314-345.

\bibitem{24} L. Friedlander, {\it On the spectrum of a class of second order
periodic elliptic differential operators}, Commun. Math. Phys. {\bf 229} (2002),
49-55. 

\bibitem{25} T. A. Suslina and R. G. Shterenberg, {\it Absolute continuity of the
spectrum of the Schr\"{o}dinger operator with the potential concentrated on a
periodic system of hypersurfaces}, Algebra i Analiz {\bf 13} (2001), no. 5, 197-240; 
English transl., St. Petersburg Math. J. {\bf 13} (2002), no. 5, 859-891. 

\bibitem{26} H. L. Cycon, R. G. Froese, W. Kirsch, and B. Simon, {\it Schr\"{o}dinger 
operators with application to quantum mechanics and global geometry}, Springer-Verlag, 
Berlin, 1987.

\bibitem{27} A. Morame, {\it Absence of singular spectrum for a perturbation of a
two-dimensional Laplace-Beltrami operator with periodic electro-magnetic
potential}, J. Phys. A: Math. Gen. {\bf 31} (1998), 7593-7601. 

\bibitem{28} L. I. Danilov, {\it On absolute continuity of the spectrum of periodic 
Schr\" odinger and Dirac operators. II}, Fiz.-Tekhn. Inst. Ural. Otdel. Ross. 
Akad. Nauk, Izhevsk, 2001. Manuscript dep. VINITI 09.04.01, no. 916-B2001 (in Russian).

\bibitem{29} L. I. Danilov, {\it On the spectrum of two-dimensional periodic
Schr\" odinger and Dirac operators}, Izv. Inst. Mat. i Inform. Udmurt. Univ., vyp.~3 
(26), Izhevsk, 2002, pp. 3-98 (in Russian).

\bibitem{30} L. I. Danilov, {\it On the spectrum of a two-dimensional periodic
Schr\" odinger operator}, Teoret. Mat. Fiz. {\bf 134} (2003), no. 3, 447-459 (in
Russian). 

\bibitem{31} L. I. Danilov, {\it On absolute continuity of the spectrum of periodic 
Schr\" odinger and Dirac operators. III}, Fiz.-Tekhn. Inst. Ural. Otdel. Ross. Akad. 
Nauk, Izhevsk, 2002. Manuscript dep. VINITI 22.10.02, no. 1798-B2002 (in Russian).

\bibitem{32} L. I. Danilov, {\it On the absence of eigenvalues in the spectrum of
two-dimensional periodic Dirac and Schr\" odinger operators}, Izv. Inst. Mat. i Inform. 
Udmurt. Univ., vyp.~1 (29), Izhevsk, 2004, pp. 49-84 (in Russian).

\bibitem{33} M. Sh. Birman, T. A. Suslina, and R. G. Shterenberg, {\it Absolute 
continuity of the spectrum of a two-dimensional Schr\"{o}dinger operator with 
potential supported on a periodic system of curves}, Algebra i Analiz {\bf 12} (2000), 
no. 6, 140-177; English transl., St. Petersburg Math. J. {\bf 12} (2001), no. 6, 
983-1012. 

\bibitem{34} R. G. Shterenberg, {\it Absolute continuity of a two-dimensional 
magnetic periodic Schr\"{o}dinger operator with electric potential of measure
derivative type}, Zap. Nauchn. Sem. S.-Peterburg. Otdel. Mat. Inst. Steklov. (POMI) 
{\bf 271} (2000), 276-312; English transl., J. Math. Sci. (New York) {\bf 115} (2003), 
no. 6, 2862-2882.

\bibitem{35} R. G. Shterenberg, {\it Absolute continuity of the spectrum of 
two-dimensional periodic Schr\"{o}dinger operators with positive electric potential}, 
Algebra i Analiz {\bf 13} (2001), no. 4, 196-228; English transl., St. Petersburg 
Math. J. {\bf 13} (2002), no. 4, 659-683. 

\bibitem{36} R. G. Shterenberg, {\it Absolute continuity of the spectrum of a
two-dimensional magnetic periodic Schr\"{o}dinger operator with positive electric 
potential}, Trudy S.Peterburg. Mat. Obshch. {\bf 9} (2001), 199-233; English transl.
in Amer. Math. Soc. Transl. Ser. 2, vol. 209, Amer. Math. Soc., Providence, RI, 2003. 

\bibitem{37} R. Schterenberg, {\it Absolute continuity of spectra of
two-dimensional periodic Schr\"{o}dinger operator with strongly subordinate
magnetic potentials}, Report no. 21, 2002/2003, Mittag-Leffler Inst., Stockholm, 2002.

\bibitem{38} T. A. Suslina and R. G. Shterenberg, {\it Absolute continuity of the
spectrum of the magnetic Schr\"{o}dinger operator with metric in a two-dimensional
periodic waveguide}, Algebra i Analiz {\bf 14} (2002), no. 2, 159-206; English 
transl., St. Petersburg Math. J. {\bf 14} (2003), no. 2, 305-343.

\bibitem{39} R. G. Shterenberg, {\it Schr\"{o}dinger operator in a periodic waveguide
on the plane and quasi-conformal mappings}, Zap. Nauchn. Sem. S.-Peterburg. Otdel. Mat. 
Inst. Steklov. (POMI) {\bf 295} (2003), 204-243 (in Russian).

\bibitem{40} A. V. Sobolev and J. Walthoe, {\it Absolute continuity in periodic
waveguides}, Proc. London Math. Soc. (3) {\bf 85} (2002), no. 3, 714-741.

\bibitem{41} E. Shargorodsky and A. V. Sobolev, {\it Quasi-conformal mappings and 
periodic spectral problems in dimension two}, LANL Archives: math.SP/0109216 (2001).

\bibitem{42} L. Thomas, {\it Time dependent approach to scattering from 
impurities in a crystal}, Commun. Math. Phys. {\bf 33} (1973), 335-343.

\bibitem{43} I. M. Gel$^{\prime }$fand, {\it Expansion in characteristic functions of
an equation with periodic coefficients}, Dokl. Akad. Nauk SSSR {\bf 73} (1950), no. 6, 
1117-1120 (in Russian). 

\bibitem{44} M. Reed and B. Simon {\it Methods of modern mathematical physics. IV.
Analysis of operators}, Acad. Press, New York - London, 1978. 

\bibitem{45} M. Reed and B. Simon {\it Methods of modern mathematical physics. III.
Scattering theory}, Acad. Press, New York - London, 1979. 

\end{thebibliography}
\end{document}